\documentclass[aps,prb,superscriptaddress,reprint]{revtex4-2}

\usepackage{graphicx}% Include figure files
\usepackage{dcolumn}% Align table columns on decimal point
\usepackage{bm}% bold math
\usepackage{amsmath}
\begin{document}

%\preprint{APS/123-QED}

\title{Unconventional Spin-Orbit Torques from Sputtered MoTe$_2$ Films}% Force line breaks with \\

\author{Shuchen Li}
    \email{sl117@illinois.edu}
\affiliation{%
Department of Materials Science and Engineering and Materials Research Laboratory, University of Illinois Urbana-Champaign, Urbana, Illinois 61801, USA
}%

\author{Jonathan Gibbons}%
\affiliation{%
Department of Materials Science and Engineering and Materials Research Laboratory, University of Illinois Urbana-Champaign, Urbana, Illinois 61801, USA
}%
\affiliation{%
Department of Physics, University of California -- San Diego, La Jolla, California 92093, USA
}%

\author{Stasiu Chyczewski}
\affiliation{Department of Electrical and Computer Engineering, University of Illinois Urbana-Champaign, Urbana, Illinois 61801, USA}

\author{Zetai Liu}
\affiliation{Department of Electrical and Computer Engineering, University of Illinois Urbana-Champaign, Urbana, Illinois 61801, USA}

\author{Hsu-Chih Ni}
\affiliation{%
Department of Materials Science and Engineering and Materials Research Laboratory, University of Illinois Urbana-Champaign, Urbana, Illinois 61801, USA
}%

\author{Jiangchao Qian}
\affiliation{%
Department of Materials Science and Engineering and Materials Research Laboratory, University of Illinois Urbana-Champaign, Urbana, Illinois 61801, USA
}%

\author{Jian-Min Zuo}
\affiliation{%
Department of Materials Science and Engineering and Materials Research Laboratory, University of Illinois Urbana-Champaign, Urbana, Illinois 61801, USA
}%

\author{Jun-Fei Zheng}
\affiliation{Entegris Inc. Danbury, Connecticut 06810}

\author{Wenjuan Zhu}
\affiliation{Department of Electrical and Computer Engineering, University of Illinois Urbana-Champaign, Urbana, Illinois 61801, USA}

\author{Axel Hoffmann}%
\email{axelh@illinois.edu}
\affiliation{%
Department of Materials Science and Engineering and Materials Research Laboratory, University of Illinois Urbana-Champaign, Urbana, Illinois 61801, USA
}%

\begin{abstract}
Materials with strong spin-orbit coupling and low crystalline symmetry are promising for generating large unconventional spin-orbit torques (SOTs), such as in-plane field-like (FL) torques and out-of-plane damping-like (DL) torques, which can effectively manipulate and deterministcally switch an out-of-plane magnetization without the need for additional external in-plane magnetic fields. Here, we report SOTs generated by magnetron-sputtered $1T'$ MoTe$_2$/Permalloy (Py; Ni$_{80}$Fe$_{20}$)/MgO heterostructures using both spin-torque ferromagnetic resonance (ST-FMR) and second harmonic Hall measurements. We observed unconventional FL and DL torques in our samples due to spins polarized normal to the interface of MoTe$_2$ and Py layers, and studied the influence of crystallographic order and MoTe$_2$ layer thickness on the SOTs. By comparing the Raman spectra of $1T'$ MoTe$_2$ samples prepared in different ways, we found a tensile strain in sputtered MoTe$_2$ films, which might further enhance the generation of unconventional torques by reducing the symmetry of $1T'$ MoTe$_2$. 
\end{abstract}

\maketitle

%\tableofcontents

\section{\label{sec:introduction} Introduction}
Spin-orbit torques (SOT) are promising for novel information technologies, such as non-volatile magnetoresistive random-access memory \cite{brataas_current-induced_2012,9288784, 9427163} as they can efficiently manipulate magnetization dynamics by transferring angular momentum to the magnetic thin films \cite{RevModPhys.91.035004} with demonstrated low power (130~pJ) and high speed (210~ps) \cite{PhysRevB.100.184402}. When applying a charge current through the spin source material, which has large spin-orbit couplings, SOTs can be generated from spin currents and concomitant spin accumulations at material boundaries, {\em e.g.}, through spin Hall effects (SHE) in bulk materials~\cite{PhysRevLett.85.393, RevModPhys.87.1213, PhysRevLett.106.036601, doi:10.1126/science.1218197, PhysRevLett.109.096602, 6516040, zhang_research_2016} and directly from electric current generated spin accumulations, {\em e.g.}, from interfacial Rashba-Edelstein effects \cite{doi:10.1021/acs.nanolett.6b03300, PhysRevB.93.180402, mellnik_spin-transfer_2014}. Such spin source materials can be heavy metals like Pt \cite{PhysRevLett.106.036601, mihai_miron_current-driven_2010} and Ta \cite{doi:10.1126/science.1218197}, topological insulators \cite{mellnik_spin-transfer_2014}, and Weyl semimetals like WTe$_2$ \cite{macneill_control_2017, li_spin-momentum_2018}.
Lately, there has been a growing interest in exotic in-plane FL torques, denoted as $\vec{\tau}{_{FL}^{z}}$, which are proportional to $\hat{m}\times\hat{z}$, as well as out-of-plane DL torques $\vec{\tau}{_{DL}^{z}}$ proportional to $\hat{m}\times(\hat{m}\times\hat{z})$, due to $z$-polarized spins (where $z$ is the interface normal direction), which can deterministacally switch magnetizations with directions $\hat{m}$ pointing out-of-plane without an external symmetry-breaking in-plane field. However, $\vec{\tau}{_{FL}^{z}}$ and $\vec{\tau}{_{DL}^{z}}$ are generally forbidden due to symmetry restrictions, but can be generated when the mirror symmetry in the plane perpendicular to the electric current direction is broken. Recent experiments have focused on exploring exotic SOTs generated from materials with intrinsically low symmetries, like WTe$_2$/Permalloy (Py; Ni$_{80}$Fe$_{20}$) \cite{macneill_control_2017} and MnPd$_3$/CoFeB \cite{dc_observation_2023}, and from materials systems with artificial symmetry breakings, like strain-induced antisymmetry in NbSe$_2$/Py devices \cite{doi:10.1021/acs.nanolett.7b04993}. Nevertheless, many questions still remain open with respect to the exact mechanisms behind the generation of those exotic torques.
%In general, when an in-plane charge current is applied through a material with high symmetry like Pt, the charge current direction ($x$), the spin current direction ($z$; perpendicular to the interface), and the spin polarization direction ($y$), will always be orthogonal to each other, such that the spin current flowing from the spin source layer into the adjacent magnetic thin film (along $z$) will carry spins polarized in-plane and have conventional in-plane damping-like (DL) torques $\vec{\tau}{_{DL}^{y}}\propto\hat{m}\times(\hat{m}\times\hat{y})$ acting on the magnetization dynamics \cite{slonczewski_current-driven_1996}, where $\hat{m}$ is the direction of the magnetization. 

In this work, we studied the SOTs generated from magnetron-sputtered MoTe$_2$ films by using both spin-torque ferromagnetic resonance (ST-FMR) \cite{PhysRevLett.106.036601, doi:10.1126/science.1218197} and second harmonic Hall measurements \cite{PhysRevB.90.224427, PhysRevB.89.144425}. We have observed sizable unconventional FL torques due to $z$-polarized spins ($\vec{\tau}{_{FL}^{z}}$) and $x$-polarized spins ($\vec{\tau}{_{FL}^{x}}\propto\hat{m}\times\hat{x}$), and DL torques due to $z$ spins ($\vec{\tau}{_{DL}^{z}}$) in our MoTe$_2$/Py/MgO devices. To investigate the origins of these torques, we studied their thickness and current direction dependencies. Through careful crystal structure characterizations, we found our sputtered MoTe$_2$ has both preferential in- and out-of-plane alignments. We also found a strain in the MoTe$_2$ films, which could further reduce the symmetry and contribute to the presence of the exotic torques.

\section{\label{sec:level2} Sample fabrication and structural characterization}

\begin{figure}[htbp]
    \centering
    \includegraphics[width=0.48\textwidth]{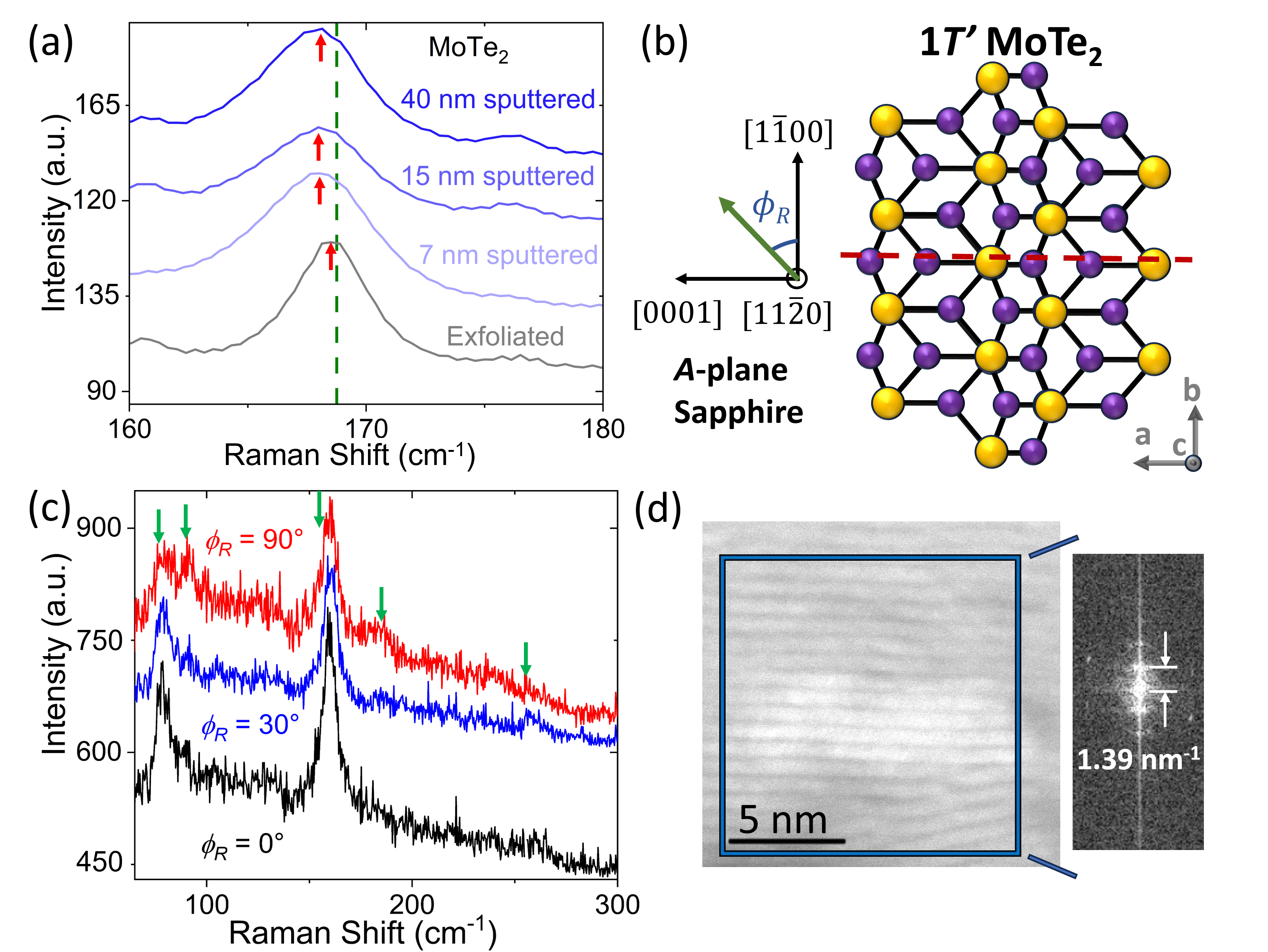}
    \caption{(a) Raman spectra of mangetron-sputtered 40-nm (dark blue), 15-nm (blue), 7-nm (light blue), and exfoliated (grey) MoTe$_2$ samples. The spectra are shifted by an offset of 10 with respect to each other. The green dashed line and the red arrows indicate the theoretical and the measured MoTe$_2$ Raman shift position. (b) Crystal structure of $1T'$ MoTe$_2$ with the only mirror plane (red dashed line) along the \textbf{a}-axis. On the left shows the $A$-plane sapphire substrate orientations. (c) Polarized Raman spectra for 15-nm MoTe$_2$/sapphire. A linearly polarized 633-nm light illuminates the sample with the polarization angle $\phi_{R}$ with respect to the [$1\bar{1}00$] direction of the $A$-plane sapphire substrate [shown in (b)]. $\phi_{R}$ = 0$^{\circ}$ means the polarization direction is parallel to the [$1\bar{1}00$] direction. The green arrows indicate different Raman modes and different peak intensities. (d) The STEM image of the sputtered MoTe$_2$ film and the fast Fourier transform of the blue circled area.} 
    \label{Raman}
\end{figure}

We synthesized MoTe$_2$ films by magnetron co-sputtering elemental Te and Mo targets onto $A$-plane sapphire (0.5-mm thick) substrates at temperatures ranging from 100--300$^\circ{C}$, followed by 1~h of annealing ranging from 350--500$^\circ{C}$ in vacuum~\cite{doi:10.1021/acsnano.0c05936}. After the samples were cooled down, we deposited 10-nm Py and 2-nm MgO capping layer in-situ before bringing the samples to ambient conditions. We performed X-ray diffraction (XRD) and scanning transmission electron microscopy (STEM) to characterize the sputtered MoTe$_2$ films. 
Fig.~\ref{Raman}(a) shows the Raman measurements on the sputtered MoTe$_2$ films and exfoliated $1T'$ MoTe$_2$ flakes (on $A$-plane sapphire), indicating the sputtered MoTe$_2$ to be $1T'$ phase, a monoclinic structure with a single mirror plane along (perpendicular to) its \textbf{a}(\textbf{b}) axis and the \textbf{c} axis to be out-of-plane [shown in Fig.~\ref{Raman}(b)]. %Furthermore, we did not observe any peak splitting \cite{zhang_raman_2016, doi:10.1021/acs.nanolett.6b02666} for the Raman mode around 130~cm$^{-1}$ that is characteristic for the $T_d$ phase MoTe$_2$.Before depositing Mo and Te, substrates were ultrasonically cleaned with aceton and subsequently IPA 2 minutes each. Thereafter, substrates were loaded into high-vacuum chamber ($5\times 10^{-9}$~Torr) and pre-heated at 500°C for 1~h to remove particles at the surface. We then co-sputtered Mo and Te at temperatures ranging from 100°C to 200°C at an Ar gas pressure of 3 mTorr, followed by 1~h of annealing at temperatures ranging from 350°C to 400°C in vacuum \cite{doi:10.1021/acsnano.0c05936}.

However, interestingly, for all MoTe$_2$ films (7~nm, 15~nm, and 40~nm) grown by magnetron sputtering, the characteristic Raman peak around 168~cm$^{-1}$ shifted to lower energies compared to the same peak from the exfoliated $1T'$ MoTe$_2$ flake. Such a red shift of the Raman peak is generally caused by a strain from a tensile stress, which will lead to the elongation of the lattice and a decrease in the bond strength between neighboring atoms~\cite{karki_strain-induced_2020, imajo_strain_2021, AngelMurriMihailovaAlvaro+2019+129+140}. We think an in-plane strain with a perpendicular component to the mirror plane of $1T'$ MoTe$_2$ was induced during high-temperature processing, such that a red shift of 168~cm$^{-1}$ Raman mode vibrating mostly in-plane perpendicular to the mirror plane of $1T'$ MoTe$_2$ \cite{PhysRevB.94.214105} was observed across the samples.
Fig.~\ref{Raman}(d) shows the STEM image of our sputtered 15-nm MoTe$_2$ sample. We calculated the distance between two adjacent bright fringes to be 1.39~nm$^{-1}$ by performing the fast Fourier transform of the blue circled area, which corresponds to 0.719 nm in real space and is close to the distance between the two MoTe$_2$ layers held by the van der Waals forces (0.693~nm), indicating the $c$-axis of MoTe$_2$ is mostly aligned vertically. 
We further explored the in-plane orientation of MoTe$_2$ films to investigate if there is any preferential alignment with the $A$-plane sapphire substrate by using polarized Raman spectroscopy. We illuminated sputtered MoTe$_2$/sapphire samples with linearly polarized light at different $\phi_{R}$, where $\phi_{R}$ is the angle between the polarization direction and the [$1\bar{1}00$] direction of the $A$-plane sapphire substrate. As shown in Fig.~\ref{Raman} (c), the polarization-dependent Raman peaks and intensities (green arrows) demonstrate there is a crystallographic texture within the sputtered MoTe$_2$, with the \textbf{a} and \textbf{b} axis of the MoTe$_2$ being preferentially aligned along the [$0001$] and [$1\bar{1}00$] directions of the $A$-plane sapphire, respectively~\cite{doi:10.1021/acsnano.6b05127,PhysRevB.100.184402}.
%Note that the characteristic Raman peak around 260~cm$^{-1}$ also has a red shift as the thickness of the MoTe$_2$ films increases from 7~nm to 40~nm. This red shift has an additional contribution from the thickness of MoTe$_2$ films. As the atomic bond of MoTe$_2$ vibrates mostly in-plane at 260~cm$^{-1}$, the average bond strength decreases as the thickness increases, which leads to a lower Raman frequency and a red shift. Such a thickness-induced red shift is also found and reported in $1T'$ MoTe$_2$ films \cite{doi:10.1021/acsnano.6b05127, keum_bandgap_2015}, whereas the red shift of the Raman peak around 168~cm$^{-1}$ has not been reported for MoTe$_2$ \cite{doi:10.1021/acsnano.6b05127, keum_bandgap_2015}, and is caused by tensile stress

\section{\label{sec:level3} Spintransport measurements} 
\subsection{Spin-Torque Ferromagnetic Resonance}
\begin{figure}[htbp]
        \centering
        \includegraphics[width=0.48\textwidth]{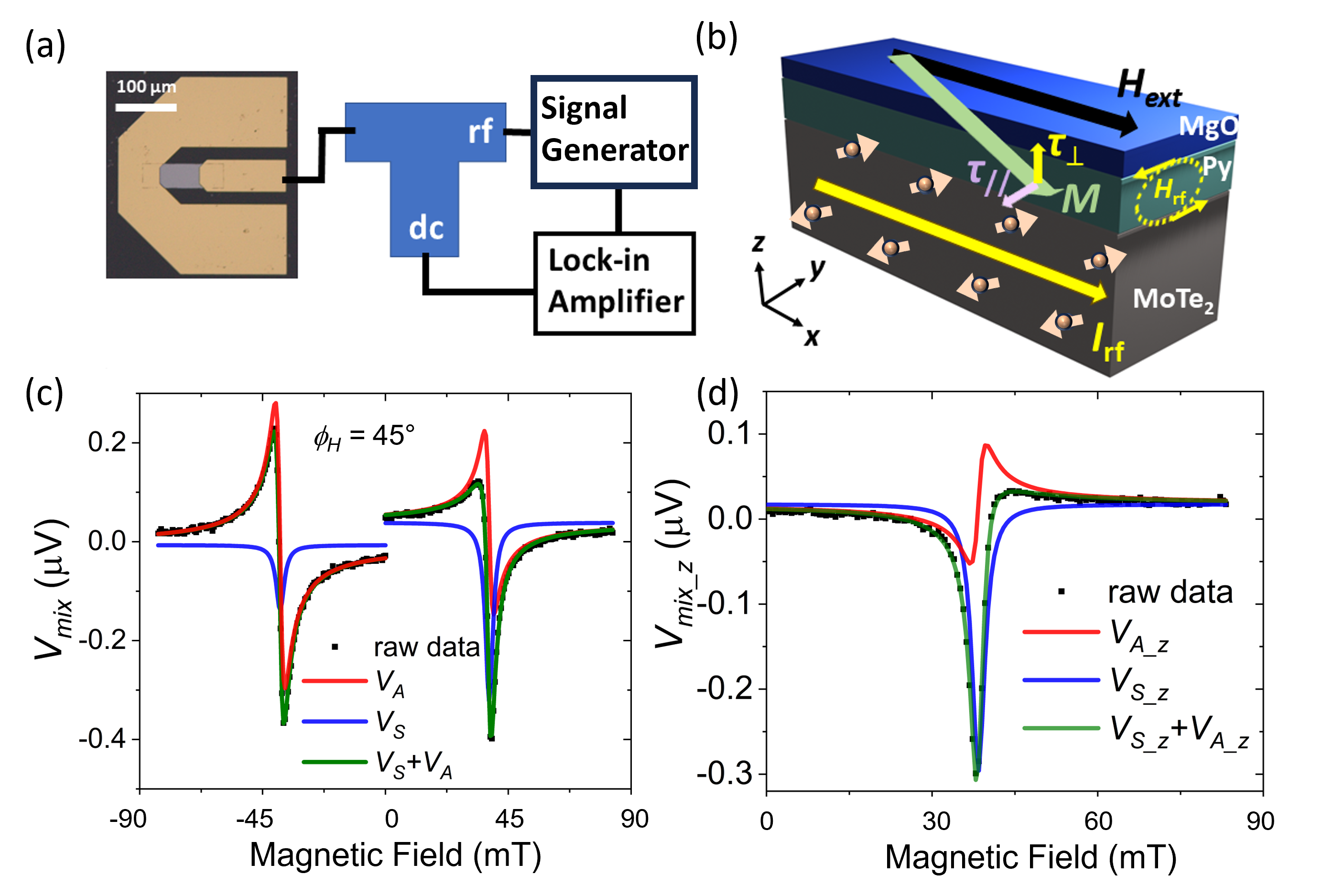}
        \caption{(a) Diagram of our measurement setup for ST-FMR. A signal generator injects a GHz {\em rf} current whose amplitude is modulated by the reference signal of a lock-in amplifier into the device through the {\em rf} port of a bias tee. The mixing {\em dc} voltage is measured by the lock-in amplifier through the {\em dc} port of the bias tee. The dimension of the device is 80--130~$\mu$m in length and 20--40~$\mu$m in width.  (b) A schematic of the spin-torque ferromagnetic resonance measurements on MoTe$_2$/Py/MgO devices. (c) The measured {\em dc} mixing voltages of sample 1 device 1 of MoTe$_2$(15)/Py/MgO at $\phi_{H}$ = 45$^\circ$ for positive and negative field scans. The power and frequency of the current is 4~dBm and 6~GHz, and the current directin is along [$1\bar{1}00$]. The fit for the mixing voltage is the green curve, which is the sum of $V_S$ (blue) and $V_A$ (red). (d) The mixing voltages $V\textsubscript{mix\_z}$ with contributions solely from $z$-polarized spins, and we found $S_z = -0.313$ and $A_z = 0.140$, which are proportional to the sizes of $\vec{\tau}{_{FL}^{z}}$ and $\vec{\tau}{_{DL}^{z}}$.}
        \label{stfmrsetup}
\end{figure}

To study the characteristics of spin-orbit torques generated from sputtered MoTe$_2$ layer, we performed spin-torque ferromagnetic resonance (ST-FMR) measurements \cite{PhysRevLett.106.036601} on our MoTe$_2$(15)/Py(10)/MgO(2) devices (the individual layer thicknesses in parenthesis are in nm). As indicated in Figs.~\ref{stfmrsetup}(a) and (b), the samples were patterned and integrated into ground-signal-ground coplanar waveguides by ion-milling and lithography, and a radio-frequency current $I_{rf}$ at 5--9~GHz along the [$1\bar{1}00$] direction was applied through our devices with an external magnetic field $H_{ext}$ sweeping in the $xy$ plane from 0.1~T to 0~T and from -0.1~T to 0~T. We then measured the resultant homodyne {\em dc} mixing voltages, $V_{mix}$, due to the coupling of the {\em rf} current and the anisotropic magnetoresistance (AMR) of Py modulated by SOTs from the MoTe$_2$ layer through a lockin amplifer. Fig.~\ref{stfmrsetup}(c) shows the measured mixing voltages $V_{mix}$ of device 1 for positive and negative field scans at $\phi_{H}$ = 45$^{\circ}$, where $\phi_{H}$ is the angle between the external magnetic field and the applied $I_{rf}$ current. $V_{mix}$ can be fitted by the sum of symmetric and antisymmetric Lorentzian functions

\begin{eqnarray}
\label{Eq:MixingVoltage}
    V_{mix} &=& S\dfrac{{\Delta}^{2}}{(H_{ext} - H_{0})^2 + {\Delta}^{2}} \nonumber \\
    & &+ A\dfrac{{\Delta}(H_{ext} - H_{0})}{(H_{ext} - H_{0})^2 + {\Delta}^{2}}~,
\end{eqnarray}

where $H_{ext}$ is the applied field, $H_0$ is the ferromagnetic resonance field of permalloy, ${\Delta}$ is the half width at half maximum resonance linewidth, and $S$ and $A$ are the fitting parameters representing the sizes of the symmetric and antisymmetric Lorentzians $V_{S}$ and $V_{A}$, which correspond to the sizes of the in-plane $\tau_{\parallel}$ and out-of-plane $\tau_{\bot}$ torques respectively. 

As shown in Fig.~\ref{stfmrsetup}(c), the different line shapes of symmetric ($V_{S}$, blue) and antisymmetric ($V_{A}$, red) Lorentzians for positive and negative field scans indicate the presence of unconventional SOTs due to $z$-polarized spins. We added the mixing voltages measured from positive field and negative field scans, which cancels voltages contributed by spins polarized in-plane, while voltages due to spins polarized out-of-plane (along $z$) will add up constructively, due to their distinct symmetry characteristics in relation to the field direction (Eqs.~\ref{Vs} and \ref{Va}). We then divided the added voltages by two and obtained the voltage ($V_{mix\_z}$), due to torques from $z$-polarized spins [Fig.~\ref{stfmrsetup}(d)]. Through fitting $V_{mix\_z}$ using Eq.~\ref{Eq:MixingVoltage}, we confirmed the existence of both $\vec{\tau}{_{DL}^{z}}$ and $\vec{\tau}{_{FL}^{z}}$ within device 1 of our MoTe$_2$(15)/Py(10)/MgO(2).

To extract different components of SOTs from $\tau_{\parallel}$ and $\tau_{\bot}$ that are related to $V_{S}$ and $V_{A}$, we varied the angle $\phi_{H}$ between the current and field, and plot the extracted $S$ and $A$ as a function of angle $\phi_{H}$. SOTs generated by $x$, $y$ and $z$-polarized spins have distinct angular dependencies for both $S$ and $A$ described by the following equations,
\begin{eqnarray}
    {S} &=& S{_{DL}^{y}}sin(2\phi_{H})cos(\phi_{H})+S{_{FL}^{z}}sin(2\phi_{H}) \nonumber \\
    & &+S{_{DL}^{x}}sin(2\phi_{H})sin(\phi_{H})
    \label{Vs} 
\end{eqnarray}  
\begin{eqnarray}
    {A} &=& A{_{FL}^{y}}sin(2\phi_{H})cos(\phi_{H})+A{_{DL}^{z}}sin(2\phi_{H}) \nonumber \\
    & &+A{_{FL}^{x}}sin(2\phi_{H})sin(\phi_{H}),
    \label{Va}
\end{eqnarray}
and we can obtain the sizes of $\vec{\tau}{_{DL}^{y}}$, $\vec{\tau}{_{FL}^{z}}$, $\vec{\tau}{_{DL}^{x}}\propto\hat{m}\times(\hat{m}\times\hat{x})$ ,$\vec{\tau}{_{FL}^{y}}$, $\vec{\tau}{_{DL}^{z}}$, $\vec{\tau}{_{FL}^{x}}$ by fitting their angular dependencies using Eqs.~\ref{Vs} and \ref{Va}, which are proportional to the fit values of $S{_{DL}^{y}}$, $S{_{FL}^{z}}$, $S{_{DL}^{x}}$, $A{_{FL}^{y}}$, $A{_{DL}^{z}}$, and $A{_{FL}^{x}}$, respectively.
 
\begin{figure}[htbp]
        \centering
        \includegraphics[width=0.48\textwidth]{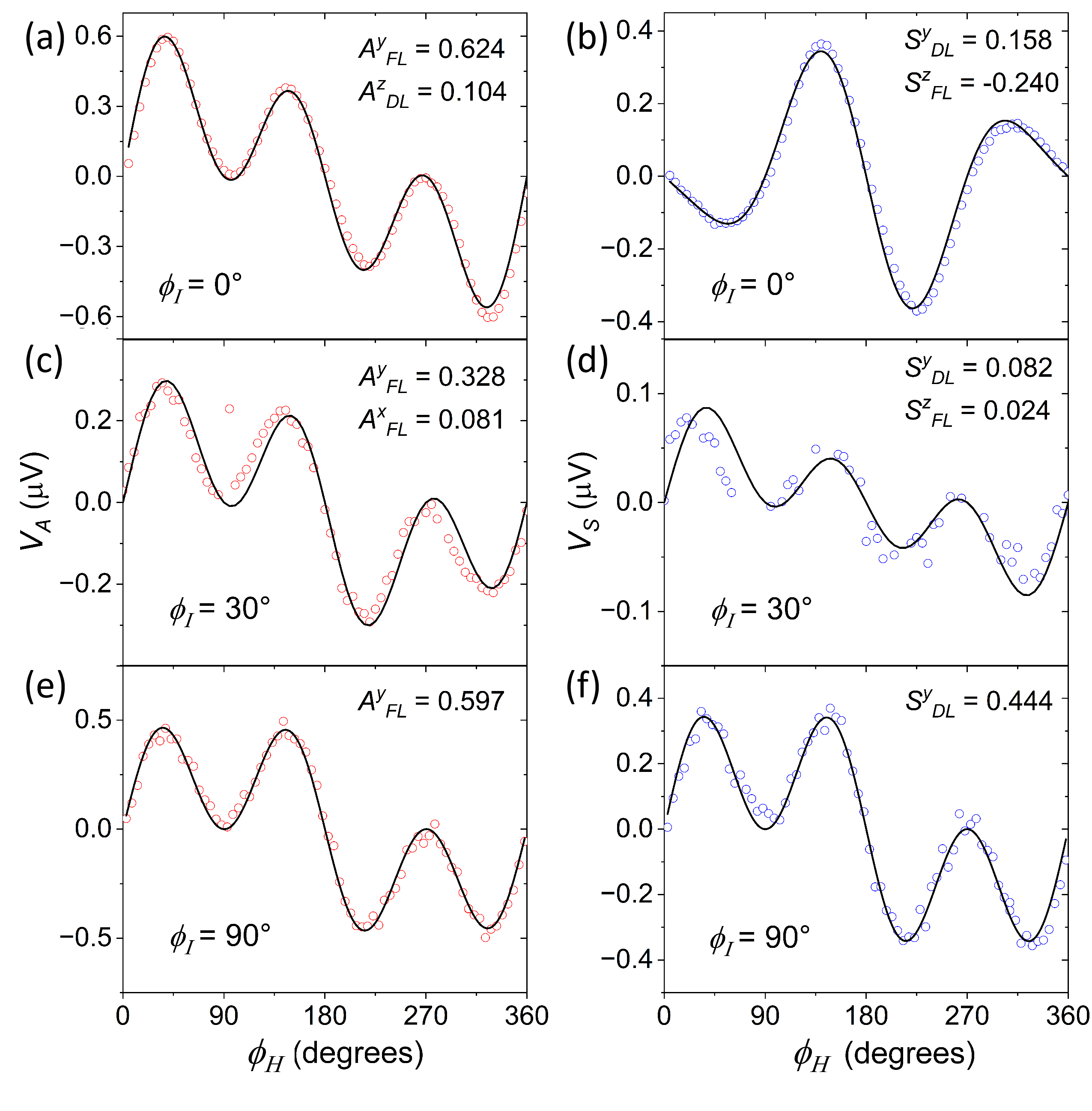}
        \caption{(a), (c) and (e) antisymmetric components $V_{A}$ as a function of angle $\phi_{H}$ for $\phi_{I}$ = 0$^\circ$, 30$^\circ$ and 90$^\circ$. (b), (d) and (f) Symmetric components $V_{S}$ as a function of angle $\phi_{H}$ for $\phi_{I}$ = 0$^\circ$, 30$^\circ$ and 90$^\circ$. The red and blue dots are extracted from the measured $V_{mix}$, and the black lines are the fitted curves using Eqs.~\ref{Vs} and \ref{Va}.}
        \label{15nmASTFMR}
\end{figure}

As shown in Figs.~\ref{15nmASTFMR}(a) and (b), in device 1 of our MoTe$_2$(15)/Py/MgO sample, we found $A{_{FL}^{y}}$ = 0.624, which we assume is mainly contributed by the Oersted field generated by the {\em rf} currents, and $S{_{DL}^{y}}$ = 0.155, related to conventional $\vec{\tau}{_{DL}^{y}}$ in our sample. Also, we found $S{_{FL}^{z}}$ = -0.240 and $A{_{DL}^{z}}$ = 0.104, indicating sizeable unconventional $\vec{\tau}{_{FL}^{z}}$ and $\vec{\tau}{_{DL}^{z}}$ due to $z$-polarized spins. In addition, we noticed that the polarity of $\vec{\tau}{_{FL}^{z}}$ is always opposite to that of $\vec{\tau}{_{DL}^{z}}$ for all the devices measured from MoTe$_2$(15)/Py/MgO. This indicates that the mechanisms behind the generation of the two torques are the same or are strongly correlated. The exact mechanism still remains unclear but this phenomenon has also been reported in the exfoliated single crystalline $1T'$ MoTe$_2$/Py samples \cite{PhysRevB.100.184402}.

\subsection{Current Direction Dependence}
We studied the effects of crystallographic order on different torque components by applying the current at different angles $\phi_{I}$ through our MoTe$_2$(15)/Py devices, where $\phi_{I}$ is the angle between the current direction $I$ and [$1\bar{1}00$] of the $A$-plane substrate, with $\phi_{I}$ = 0$^\circ$ being $I$ parallel to [$1\bar{1}00$]. As shown in Figs.~\ref{15nmASTFMR}(c) and (d), we found sizeable $\vec{\tau}{_{FL}^{x}}$ due to $x$-polarized spins in the antisymmetric component when $\phi_{I}$ = 30${^\circ}$, but no $\vec{\tau}{_{DL}^{z}}$ from the antisymmetric component. Moreover, we found the sizes of all unconventional torques diminished at $\phi_{I}$ = 90${^\circ}$, and only the conventional torques $\vec{\tau}{_{DL}^{y}}$ remained [Figs.\ref{15nmASTFMR}(e) and (f)].

\begin{figure}[htbp]
        \centering
        \includegraphics[width=0.5\textwidth]{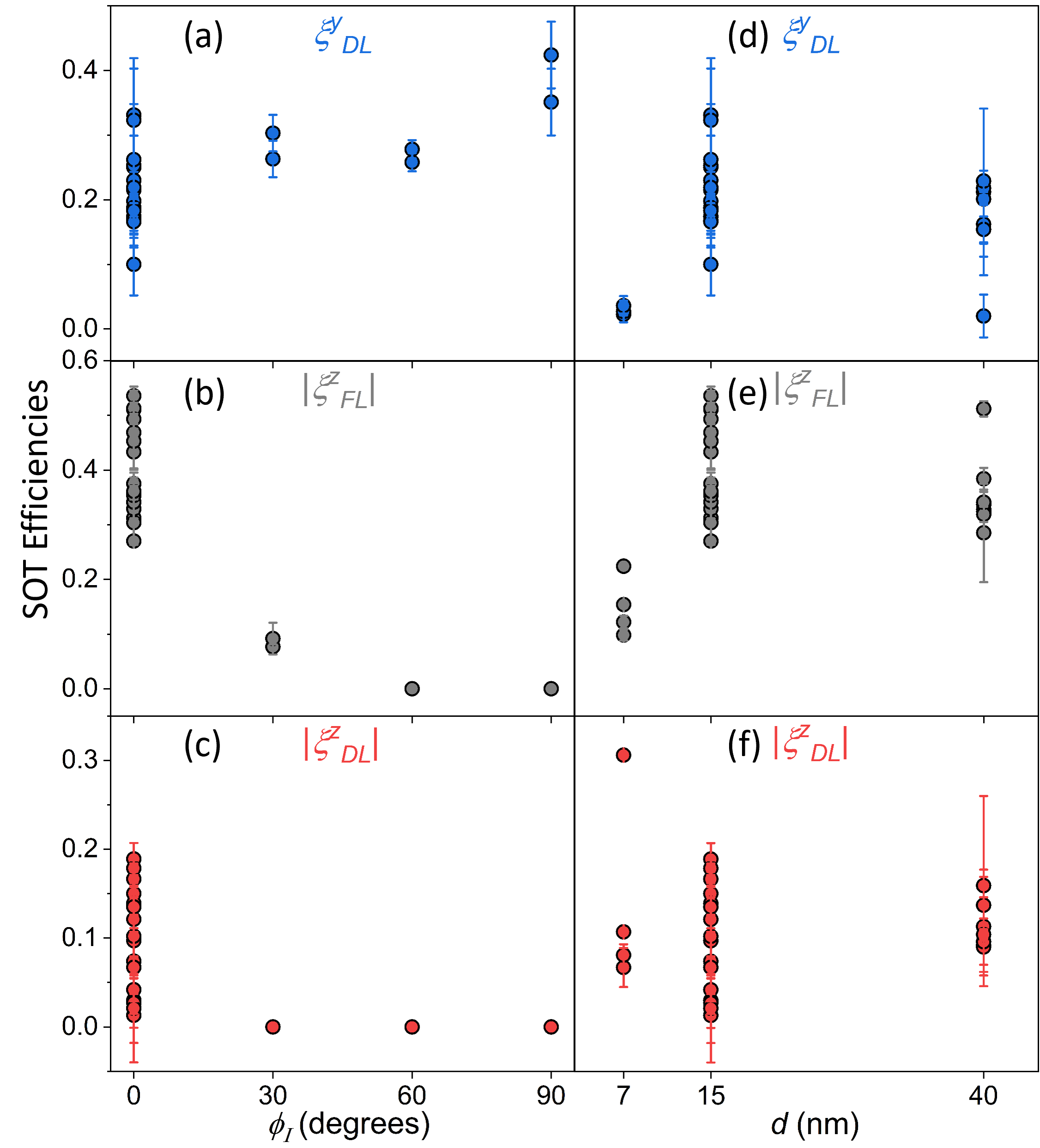}
        \caption{(a) $\xi{_{DL}^{y}}$ at different $\phi_{I}$, (b) and (c) are absolute values of $\xi{_{FL}^{z}}$ and $\xi{_{DL}^{z}}$ for better study the trend with respect to $\phi_{I}$. (d), (e) and (f) are $\xi{_{DL}^{y}}$, $|\xi{_{FL}^{z}}|$ and $|\xi{_{DL}^{z}}|$ for devices with different MoTe$_2$ thicknesses (7, 15 and 40~nm).}
        \label{SOTefficiencies}
\end{figure}

The spin-orbit torque (SOT) efficiency quantifies the ability of the spin-orbit coupling to convert an applied electric current into a torque that influences the magnetization dynamics, and we calculated the SOT efficiencies $\xi{_{DL}^{y}}$, $\xi{_{FL}^{z}}$ and $\xi{_{DL}^{z}}$ using the equation 
\begin{equation}
    \xi = \dfrac{(S{_{DL}^{y}}, {S{_{FL}^{z}}}, or {A{_{DL}^{z}}})}{A{_{FL}^{y}}}\dfrac{e\mu_{0}M_{S}td}{\hbar}[1+M_{eff}/H_{0}]^{1/2},
    \label{SOTe}
\end{equation}
where $t$ and $d$ are the thicknesses of Py and MoTe$_2$ layers, $M_S$ is the saturation magnetization of Py, which is approximately $\mu_0M_S \approx 0.8$~T. Figs.~\ref{SOTefficiencies}(a)--(c) show the calculated SOT efficiencies measured across different devices from MoTe$_2$(15)/Py/MgO heterostructures at different $\phi_{I}$.

We found $\xi{_{DL}^{y}}$ did not display a clear dependence on $\phi_{I}$, whereas $\xi{_{FL}^{z}}$ and $\xi{_{DL}^{z}}$ both showed strong relations to current directions. When the current was applied at an angle to the [$1\bar{1}00$] direction, $\xi{_{DL}^{z}}$ vanished, while $\xi{_{FL}^{z}}$ decreased at $\phi_{I}$ = 30${^\circ}$ and diminished further at $\phi_{I}$ = 60${^\circ}$ and 90${^\circ}$. For $\xi{_{DL}^{z}}$ and $\xi{_{FL}^{z}}$, the strong $\phi_{I}$ dependence indicates symmetry-related origin for the generation of $\vec{\tau}{_{DL}^{z}}$ and $\vec{\tau}{_{FL}^{z}}$, which would emerge when a current is applied perpendicular to the mirror plane of $1T'$ MoTe$_2$, and become symmetry-forbidden when the current is applied along the mirror plane.

\subsection{Thickness Dependence}
To better understand origins of different spin-orbit torque components, we investigated the thickness dependencies of different torques. Similarly, we performed ST-FMR measurements on MoTe$_2$(40 or 7)/Py/MgO heterostructures and calculated $\xi{_{DL}^{y}}$, $\xi{_{DL}^{z}}$ and $\xi{_{FL}^{z}}$ using Eq.~\ref{SOTe} for all the devices measured. As shown in Figs.~\ref{SOTefficiencies}(d)--(f), $\xi{_{DL}^{y}}$ and $\xi{_{FL}^{z}}$ increased when the MoTe$_{2}$ thickness $d$ increased from 7~nm to 15~nm, and became saturated when $d$ = 40 nm, indicating possible origns to be bulk effects. The $\phi_{I}$ independence of $\xi{_{DL}^{y}}$ suggests that $\vec{\tau}{_{DL}^{y}}$ originates from the spin Hall effect (SHE). For $\xi{\tau}{_{DL}^{z}}$, we did not observe any clear trend with respect to $d$ [Fig.~\ref{SOTefficiencies}(f)], which would imply the origin of $\vec{\tau}{_{DL}^{z}}$ to be related to the interfacial properties between MoTe$_{2}$ and Py. Finally, for MoTe$_2$(40) and MoTe$_2$(7), we found the polarities of $\xi{_{FL}^{z}}$ and $\xi{_{DL}^{z}}$ were always opposite for all the devices measured, just as what has been found in MoTe$_2$(15).

\subsection{Second Harmonic Hall Measurements}

\begin{figure}[htbp]
        \centering
        \includegraphics[width=\columnwidth]{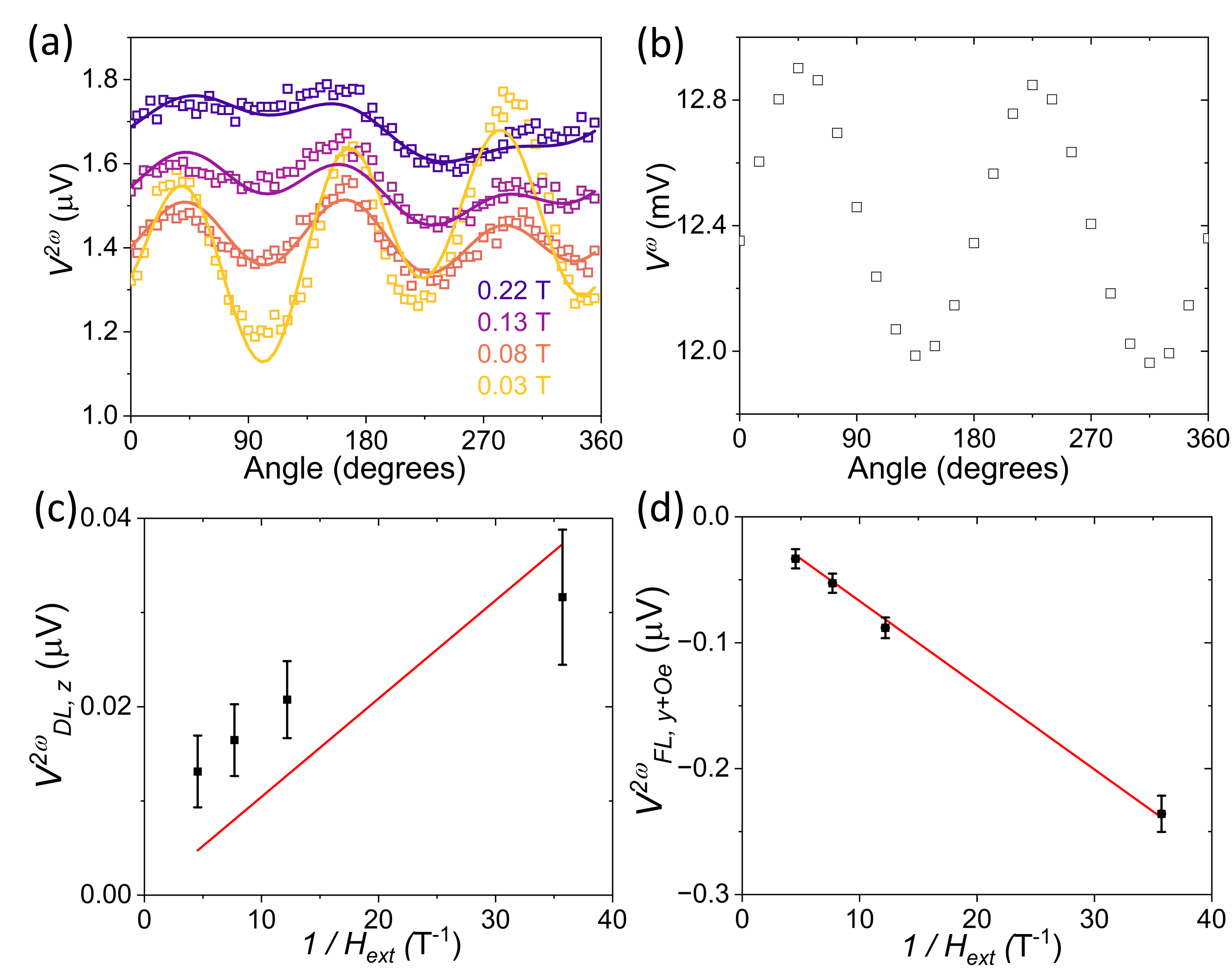}
        \caption{(a) $V^{2\omega}$ (dots) of MoTe$_2$(15)/Py/MgO as a function of $\phi_{H}$ for various fields and the fit curves (lines) using Eq.~\ref{secondharmonic}. (b) $V^{\omega}$ as a function of $\phi_{H}$ under $H_{ext}$ = 0.22 T. (c) and (d) Components of $V^{2\omega}$, $V^{2\omega}_{DL,z}$ and $V^{2\omega}_{FL,y + Oe}$, contributed by $H{_{DL}^{z}}$ and $H{_{FL}^{y}} + H_{Oe}$, with linear fit red lines.}
        \label{SHH}
\end{figure}

Second harmonic Hall measurements are useful for analyzing in-plane $\tau_{\parallel}$ and out-of-plane $\tau_{\bot}$ spin-orbit torques, and can be complimentary to ST-FMR. For our MoTe$_2$(15)/Py/MgO sample, we applied a low frequency ($\omega$ = 1131~kHz) {\em ac} current with 4 mA amplitude along the $x$ direction, with the magnetization of Py defined in-plane by an external magnetic field ($H_{ext}$), and measured the transverse second harmonic Hall voltage $V^{2\omega}$ along $y$. $\tau_{\parallel}$($\tau_{\bot}$) generated by MoTe$_2$ has an effective out-of-plane(in-plane) field that rotates the magnetization of Py out-of-plane(in-plane) such that it modulates the change of the anomalous Hall resistance $R_{AHE}$ (planar Hall resistance $R_{PHE}$) at a frequency $\omega$ of the {\em ac} current. The change of $R_{AHE}$ and $R_{PHE}$ couples to the applied current and gives rise to Hall voltages $V^{2\omega}$ on the second harmonic $2\omega$ that can be detected through a lock-in amplifier. Figure~\ref{SHH}(a) shows $V^{2\omega}$ as a function of $\phi_{H}$, where $\phi_{H}$ is the angle between $H_{ext}$ and the applied current, at various fields. For an in-plane magnetization system, different spin-orbit torque components can be extracted by their distinct angular dependencies described in the following equation~\cite{PhysRevB.89.144425, PhysRevLett.104.046601, macneill_control_2017}
\begin{widetext}
\begin{eqnarray}
    V^{2\omega} &=& V_{PHE}cos(2\phi_{H})(\dfrac{H{_{FL}^{y}}+ H_{Oe}}{H_{ext}}cos(\phi_{H}) + \dfrac{H{_{FL}^{x}}}{H_{ext}}sin(\phi_{H}) +\dfrac{H{_{DL}^{z}}}{H_{ext}})\nonumber \\ 
    & &+ \dfrac{1}{2}V_{AHE}(\dfrac{H{_{DL}^{y}}}{H_{ext}+H_{k}}cos(\phi_{H})
    + \dfrac{H{_{DL}^{x}}}{H_{ext}+H_{k}}sin(\phi_{H})
    + \dfrac{H{_{FL}^{z}}}{H_{ext}+H_{k}}) + C~,
    \label{secondharmonic}
\end{eqnarray}
\end{widetext}
where $H{_{FL}^{y}}$, $H{_{FL}^{x}}$, $H{_{DL}^{z}}$, $H{_{DL}^{y}}$, $H{_{DL}^{x}}$ and $H{_{FL}^{z}}$ are the spin-orbit fields corresponding to the respective torques, and $C$ is an offset constant. Through fitting, we observed sizable contributions from $\vec{\tau}{_{FL}^{y}}$ + $\vec{\tau}_{Oe}$, $\vec{\tau}{_{DL}^{y}}$ and $\vec{\tau}{_{DL}^{z}}$ to $V^{2\omega}$, and plot the extracted components $V^{2\omega}_{DL,z}$ and $V^{2\omega}_{FL,y + Oe}$ as a function of $1/H_{ext}$ [see Figs.~\ref{SHH}(c) and (d)]. Figure~\ref{SHH}(b) shows the angular dependencies of the transverse $V^{\omega}$, from which we derived the planar Hall effect ($V_{PHE}$) to be 0.45 mV. Thus, we calculated $H{_{FL}^{y}} + H_{Oe}$ and $H{_{DL}^{z}}$ to be 8.0 A/m and 1.67 A/m. 
Finally, we calculated the SOT efficiency $\xi^{z}_{DL}$ = 0.068 using
\begin{equation}
    \xi^{z}_{DL} = \dfrac{2eM_{s}t_{FM}}{\hbar} \dfrac{H^{z}_{DL}}{J_{MoTe_2}},
    \label{SHHSOTefficiency}
\end{equation}
where $e$, $t$ and $\hbar$ are the electron charge, thickness of permalloy and reduced Planck's constant. $J_{MoTe_2} = I_{MoTe_2}/(dw)$ is the current density flowing through the MoTe$_2$ layer, where $I_{MoTe_2}$ is estimated to be 0.27~mA (see Supplemental Material S1), $d$ = 15 nm and $w$  = 20~$\mu{m}$ is the width of the device. We also estimated the Oersted field generated by $I_{MoTe_2}$ to be 6.75 A/m using Ampere’s law $H_{Oe} = I_{MoTe_2}/2w$ assuming the sample to be an infinitely wide plate.  This confirms that the Oersted field dominates $V^{2\omega}_{FL,y + Oe}$.

\section{\label{sec:level4} Discussion}
The observations of unconventional spin-orbit torques in magnetron-sputtered $1T'$ MoTe$_2$ films are possibly caused by two effects. First, as it can seen from Figs.~\ref{Raman}(c) and (d), the sputtered MoTe$_2$ films have a textured structure with the \textbf{c}-axis mostly aligned vertically and \textbf{a}(\textbf{b}) axis preferentially aligned along the [$0001$]([$1\bar{1}00$]) direction of the $A$-plane sapphire substrate, such that it preserves some macroscopic low symmetries that lack inversion symmetry and has only one mirror plane, enabling the generation of symmetry-forbidden SOTs when a current is applied at an angle to the mirror plane breaking this symmetry. 
%However, the polycrystalline nature of the MoTe$_2$ films did cause a variability of the torque efficiencies across devices with the same thickness, but an overall discernible trend with the thickness was also observed for $\vec{\tau}{_{DL}^{y}}$ and $\vec{\tau}{_{FL}^{z}}$.

Second, a strain identified from the red shift of the 168~cm$^{-1}$ Raman mode for all sputtered MoTe$_2$ films can further reduce the symmetry and enhance the effects generating unconventional spin-orbit torques. Such a strain can be caused by a tensile stress introduced by the mismatch between the thermal expansion coefficients of the substrates \cite{tian_temperature-dependent_2016, doi:10.1021/nl201488g} and MoTe$_2$ during high temperature growth. It is possible that the large SOT efficiency range in $1T'$ MoTe$_2$ films is caused by the relative alignment of the strain axis with the local grain orientations, analogous to the strain-induced $\vec{\tau}{_{FL}^{z}}$ found in single crystalline NbSe$_2$ \cite{doi:10.1021/acs.nanolett.7b04993}, and $\vec{\tau}{_{DL}^{z}}$ in exfoliated single crystalline $1T'$ MoTe$_2$ \cite{PhysRevB.100.184402}.

Finally, we performed similar measurements on $1T'$ MoTe$_2$ grown by metal-organic chemical vapor deposition\cite{2404.19135}, and observed different results [see S4 in Supplemental Materials].
%In this experiment, we observed both $\vec{\tau}{_{FL}^{z}}$ and $\vec{\tau}{_{DL}^{z}}$ in sputtered bulk poly-crystalline MoTe$_2$ samples with varying thicknesses. The torque conductivities $\sigma_{_{FL}^{z}} [50\times10^{3}\hbar/2e(\Omega^{-1}{m}^{-1})]$ and $\sigma_{_{DL}^{z}} 
%[10\times10^{3}\hbar/2e(\Omega^{-1}{m}^{-1})]$ are an order of magnitude larger than those reported in NbSe$_{2}$ $[\sigma_{T}\approx2\times10^{3}\hbar/2e(\Omega^{-1}{m}^{-1})]$ and exfoliated MoTe$_2$[$\sigma_{B}\approx1.5\times10^{3}\hbar/2e(\Omega^{-1}{m}^{-1})$].

\section{\label{sec:level5} Conclusion}
In conclusion, we have studied and reported spin-torques found in magnetron-sputtered MoTe$_2$/Py devices by performing ST-FMR and second harmonic Hall measurements on MoTe$_2$/Py(10)/MgO(2) samples. We explored the origins of different SOTs by studying the influence of the crystallographic order and the MoTe$_2$ thickness on the SOT efficiencies. Based on the $\phi_{I}$ and thickness $d$ dependence of SOT efficiencies of different torques, the origins for $\vec{\tau}{_{DL}^{y}}$ and $\vec{\tau}{_{FL}^{z}}$ are probably due to bulk spin-orbit coupling, specifically the spin Hall effect for $\vec{\tau}{_{DL}^{y}}$. For $\vec{\tau}{_{DL}^{z}}$, the origins may be related to interfacial properties between the sputtered MoTe$_2$ films and the magnetic layer.  Furthermore, the origins of $\vec{\tau}{_{FL}^{z}}$ and $\vec{\tau}{_{DL}^{z}}$ are closely related as their polarities are always opposite to each other. Our findings show that strains can be promising for generating large exotic in-plane FL and out-of-plane DL torques by introducing additional symmetry reduction, and single crystallinity are not necessary in generating exotic torques.

\section{Acknowledgement}
This work was supported by the National Science Foundation under Grant No. ECCS-2031870.  In addition data analysis and discussion by J.G. was supported as part of Quantum Materials for Energy Efficient Neuromorphic Computing (Q-MEEN-C), an Energy Frontier Research Center funded by the US Department of Energy (DOE), Office of Science, Basic Energy Sciences (BES), under Award No. DE-SC0019273. S.C., Z.L., and W.Z. would like to acknowledge the support from Entegris Incorporated under grant No. Entegris 108252.

% \nocite{*}

\bibliography{MoTe2}% Produces the bibliography via BibTeX.

\providecommand{\noopsort}[1]{}\providecommand{\singleletter}[1]{#1}%
\begin{thebibliography}{32}%
\makeatletter
\providecommand \@ifxundefined [1]{%
 \@ifx{#1\undefined}
}%
\providecommand \@ifnum [1]{%
 \ifnum #1\expandafter \@firstoftwo
 \else \expandafter \@secondoftwo
 \fi
}%
\providecommand \@ifx [1]{%
 \ifx #1\expandafter \@firstoftwo
 \else \expandafter \@secondoftwo
 \fi
}%
\providecommand \natexlab [1]{#1}%
\providecommand \enquote  [1]{``#1''}%
\providecommand \bibnamefont  [1]{#1}%
\providecommand \bibfnamefont [1]{#1}%
\providecommand \citenamefont [1]{#1}%
\providecommand \href@noop [0]{\@secondoftwo}%
\providecommand \href [0]{\begingroup \@sanitize@url \@href}%
\providecommand \@href[1]{\@@startlink{#1}\@@href}%
\providecommand \@@href[1]{\endgroup#1\@@endlink}%
\providecommand \@sanitize@url [0]{\catcode `\\12\catcode `\$12\catcode
  `\&12\catcode `\#12\catcode `\^12\catcode `\_12\catcode `\%12\relax}%
\providecommand \@@startlink[1]{}%
\providecommand \@@endlink[0]{}%
\providecommand \url  [0]{\begingroup\@sanitize@url \@url }%
\providecommand \@url [1]{\endgroup\@href {#1}{\urlprefix }}%
\providecommand \urlprefix  [0]{URL }%
\providecommand \Eprint [0]{\href }%
\providecommand \doibase [0]{https://doi.org/}%
\providecommand \selectlanguage [0]{\@gobble}%
\providecommand \bibinfo  [0]{\@secondoftwo}%
\providecommand \bibfield  [0]{\@secondoftwo}%
\providecommand \translation [1]{[#1]}%
\providecommand \BibitemOpen [0]{}%
\providecommand \bibitemStop [0]{}%
\providecommand \bibitemNoStop [0]{.\EOS\space}%
\providecommand \EOS [0]{\spacefactor3000\relax}%
\providecommand \BibitemShut  [1]{\csname bibitem#1\endcsname}%
\let\auto@bib@innerbib\@empty
%</preamble>
\bibitem [{\citenamefont {Brataas}\ \emph {et~al.}(2012)\citenamefont
  {Brataas}, \citenamefont {Kent},\ and\ \citenamefont
  {Ohno}}]{brataas_current-induced_2012}%
  \BibitemOpen
  \bibfield  {author} {\bibinfo {author} {\bibfnamefont {A.}~\bibnamefont
  {Brataas}}, \bibinfo {author} {\bibfnamefont {A.~D.}\ \bibnamefont {Kent}},\
  and\ \bibinfo {author} {\bibfnamefont {H.}~\bibnamefont {Ohno}},\ }\bibfield
  {title} {\bibinfo {title} {Current-induced torques in magnetic materials},\
  }\href {https://doi.org/10.1038/nmat3311} {\bibfield  {journal} {\bibinfo
  {journal} {Nat. Mater.}\ }\textbf {\bibinfo {volume} {11}},\ \bibinfo {pages}
  {372} (\bibinfo {year} {2012})}\BibitemShut {NoStop}%
\bibitem [{\citenamefont {Natsui}\ \emph {et~al.}(2021)\citenamefont {Natsui},
  \citenamefont {Tamakoshi}, \citenamefont {Honjo}, \citenamefont {Watanabe},
  \citenamefont {Nasuno}, \citenamefont {Zhang}, \citenamefont {Tanigawa},
  \citenamefont {Inoue}, \citenamefont {Niwa}, \citenamefont {Yoshiduka},
  \citenamefont {Noguchi}, \citenamefont {Yasuhira}, \citenamefont {Ma},
  \citenamefont {Shen}, \citenamefont {Fukami}, \citenamefont {Sato},
  \citenamefont {Ikeda}, \citenamefont {Ohno}, \citenamefont {Endoh},\ and\
  \citenamefont {Hanyu}}]{9288784}%
  \BibitemOpen
  \bibfield  {author} {\bibinfo {author} {\bibfnamefont {M.}~\bibnamefont
  {Natsui}}, \bibinfo {author} {\bibfnamefont {A.}~\bibnamefont {Tamakoshi}},
  \bibinfo {author} {\bibfnamefont {H.}~\bibnamefont {Honjo}}, \bibinfo
  {author} {\bibfnamefont {T.}~\bibnamefont {Watanabe}}, \bibinfo {author}
  {\bibfnamefont {T.}~\bibnamefont {Nasuno}}, \bibinfo {author} {\bibfnamefont
  {C.}~\bibnamefont {Zhang}}, \bibinfo {author} {\bibfnamefont
  {T.}~\bibnamefont {Tanigawa}}, \bibinfo {author} {\bibfnamefont
  {H.}~\bibnamefont {Inoue}}, \bibinfo {author} {\bibfnamefont
  {M.}~\bibnamefont {Niwa}}, \bibinfo {author} {\bibfnamefont {T.}~\bibnamefont
  {Yoshiduka}}, \bibinfo {author} {\bibfnamefont {Y.}~\bibnamefont {Noguchi}},
  \bibinfo {author} {\bibfnamefont {M.}~\bibnamefont {Yasuhira}}, \bibinfo
  {author} {\bibfnamefont {Y.}~\bibnamefont {Ma}}, \bibinfo {author}
  {\bibfnamefont {H.}~\bibnamefont {Shen}}, \bibinfo {author} {\bibfnamefont
  {S.}~\bibnamefont {Fukami}}, \bibinfo {author} {\bibfnamefont
  {H.}~\bibnamefont {Sato}}, \bibinfo {author} {\bibfnamefont {S.}~\bibnamefont
  {Ikeda}}, \bibinfo {author} {\bibfnamefont {H.}~\bibnamefont {Ohno}},
  \bibinfo {author} {\bibfnamefont {T.}~\bibnamefont {Endoh}},\ and\ \bibinfo
  {author} {\bibfnamefont {T.}~\bibnamefont {Hanyu}},\ }\bibfield  {title}
  {\bibinfo {title} {Dual-port {SOT-MRAM} achieving {90-MHz} read and {60-MHz}
  write operations under field-assistance-free condition},\ }\href
  {https://doi.org/10.1109/JSSC.2020.3039800} {\bibfield  {journal} {\bibinfo
  {journal} {IEEE Journal of Solid-State Circuits}\ }\textbf {\bibinfo {volume}
  {56}},\ \bibinfo {pages} {1116} (\bibinfo {year} {2021})}\BibitemShut
  {NoStop}%
\bibitem [{\citenamefont {Shao}\ \emph {et~al.}(2021)\citenamefont {Shao},
  \citenamefont {Li}, \citenamefont {Liu}, \citenamefont {Yang}, \citenamefont
  {Fukami}, \citenamefont {Razavi}, \citenamefont {Wu}, \citenamefont {Wang},
  \citenamefont {Freimuth}, \citenamefont {Mokrousov}, \citenamefont {Stiles},
  \citenamefont {Emori}, \citenamefont {Hoffmann}, \citenamefont {Åkerman},
  \citenamefont {Roy}, \citenamefont {Wang}, \citenamefont {Yang},
  \citenamefont {Garello},\ and\ \citenamefont {Zhang}}]{9427163}%
  \BibitemOpen
  \bibfield  {author} {\bibinfo {author} {\bibfnamefont {Q.}~\bibnamefont
  {Shao}}, \bibinfo {author} {\bibfnamefont {P.}~\bibnamefont {Li}}, \bibinfo
  {author} {\bibfnamefont {L.}~\bibnamefont {Liu}}, \bibinfo {author}
  {\bibfnamefont {H.}~\bibnamefont {Yang}}, \bibinfo {author} {\bibfnamefont
  {S.}~\bibnamefont {Fukami}}, \bibinfo {author} {\bibfnamefont
  {A.}~\bibnamefont {Razavi}}, \bibinfo {author} {\bibfnamefont
  {H.}~\bibnamefont {Wu}}, \bibinfo {author} {\bibfnamefont {K.}~\bibnamefont
  {Wang}}, \bibinfo {author} {\bibfnamefont {F.}~\bibnamefont {Freimuth}},
  \bibinfo {author} {\bibfnamefont {Y.}~\bibnamefont {Mokrousov}}, \bibinfo
  {author} {\bibfnamefont {M.~D.}\ \bibnamefont {Stiles}}, \bibinfo {author}
  {\bibfnamefont {S.}~\bibnamefont {Emori}}, \bibinfo {author} {\bibfnamefont
  {A.}~\bibnamefont {Hoffmann}}, \bibinfo {author} {\bibfnamefont
  {J.}~\bibnamefont {Åkerman}}, \bibinfo {author} {\bibfnamefont
  {K.}~\bibnamefont {Roy}}, \bibinfo {author} {\bibfnamefont {J.-P.}\
  \bibnamefont {Wang}}, \bibinfo {author} {\bibfnamefont {S.-H.}\ \bibnamefont
  {Yang}}, \bibinfo {author} {\bibfnamefont {K.}~\bibnamefont {Garello}},\ and\
  \bibinfo {author} {\bibfnamefont {W.}~\bibnamefont {Zhang}},\ }\bibfield
  {title} {\bibinfo {title} {Roadmap of spin–orbit torques},\ }\href
  {https://doi.org/10.1109/TMAG.2021.3078583} {\bibfield  {journal} {\bibinfo
  {journal} {IEEE Trans. Magn,}\ }\textbf {\bibinfo {volume} {57}},\ \bibinfo
  {pages} {1} (\bibinfo {year} {2021})}\BibitemShut {NoStop}%
\bibitem [{\citenamefont {Manchon}\ \emph {et~al.}(2019)\citenamefont
  {Manchon}, \citenamefont {\ifmmode~\check{Z}\else \v{Z}\fi{}elezn\'y},
  \citenamefont {Miron}, \citenamefont {Jungwirth}, \citenamefont {Sinova},
  \citenamefont {Thiaville}, \citenamefont {Garello},\ and\ \citenamefont
  {Gambardella}}]{RevModPhys.91.035004}%
  \BibitemOpen
  \bibfield  {author} {\bibinfo {author} {\bibfnamefont {A.}~\bibnamefont
  {Manchon}}, \bibinfo {author} {\bibfnamefont {J.}~\bibnamefont
  {\ifmmode~\check{Z}\else \v{Z}\fi{}elezn\'y}}, \bibinfo {author}
  {\bibfnamefont {I.~M.}\ \bibnamefont {Miron}}, \bibinfo {author}
  {\bibfnamefont {T.}~\bibnamefont {Jungwirth}}, \bibinfo {author}
  {\bibfnamefont {J.}~\bibnamefont {Sinova}}, \bibinfo {author} {\bibfnamefont
  {A.}~\bibnamefont {Thiaville}}, \bibinfo {author} {\bibfnamefont
  {K.}~\bibnamefont {Garello}},\ and\ \bibinfo {author} {\bibfnamefont
  {P.}~\bibnamefont {Gambardella}},\ }\bibfield  {title} {\bibinfo {title}
  {Current-induced spin-orbit torques in ferromagnetic and antiferromagnetic
  systems},\ }\href {https://doi.org/10.1103/RevModPhys.91.035004} {\bibfield
  {journal} {\bibinfo  {journal} {Rev. Mod. Phys.}\ }\textbf {\bibinfo {volume}
  {91}},\ \bibinfo {pages} {035004} (\bibinfo {year} {2019})}\BibitemShut
  {NoStop}%
\bibitem [{\citenamefont {Stiehl}\ \emph {et~al.}(2019)\citenamefont {Stiehl},
  \citenamefont {Li}, \citenamefont {Gupta}, \citenamefont {Baggari},
  \citenamefont {Jiang}, \citenamefont {Xie}, \citenamefont {Kourkoutis},
  \citenamefont {Mak}, \citenamefont {Shan}, \citenamefont {Buhrman},\ and\
  \citenamefont {Ralph}}]{PhysRevB.100.184402}%
  \BibitemOpen
  \bibfield  {author} {\bibinfo {author} {\bibfnamefont {G.~M.}\ \bibnamefont
  {Stiehl}}, \bibinfo {author} {\bibfnamefont {R.}~\bibnamefont {Li}}, \bibinfo
  {author} {\bibfnamefont {V.}~\bibnamefont {Gupta}}, \bibinfo {author}
  {\bibfnamefont {I.~E.}\ \bibnamefont {Baggari}}, \bibinfo {author}
  {\bibfnamefont {S.}~\bibnamefont {Jiang}}, \bibinfo {author} {\bibfnamefont
  {H.}~\bibnamefont {Xie}}, \bibinfo {author} {\bibfnamefont {L.~F.}\
  \bibnamefont {Kourkoutis}}, \bibinfo {author} {\bibfnamefont {K.~F.}\
  \bibnamefont {Mak}}, \bibinfo {author} {\bibfnamefont {J.}~\bibnamefont
  {Shan}}, \bibinfo {author} {\bibfnamefont {R.~A.}\ \bibnamefont {Buhrman}},\
  and\ \bibinfo {author} {\bibfnamefont {D.~C.}\ \bibnamefont {Ralph}},\
  }\bibfield  {title} {\bibinfo {title} {Layer-dependent spin-orbit torques
  generated by the centrosymmetric transition metal dichalcogenide
  {$\ensuremath{\beta}\ensuremath{-}{\mathrm{MoTe}}_{2}$}},\ }\href
  {https://doi.org/10.1103/PhysRevB.100.184402} {\bibfield  {journal} {\bibinfo
   {journal} {Phys. Rev. B}\ }\textbf {\bibinfo {volume} {100}},\ \bibinfo
  {pages} {184402} (\bibinfo {year} {2019})}\BibitemShut {NoStop}%
\bibitem [{\citenamefont {Zhang}(2000)}]{PhysRevLett.85.393}%
  \BibitemOpen
  \bibfield  {author} {\bibinfo {author} {\bibfnamefont {S.}~\bibnamefont
  {Zhang}},\ }\bibfield  {title} {\bibinfo {title} {Spin {Hall} effect in the
  presence of spin diffusion},\ }\href
  {https://doi.org/10.1103/PhysRevLett.85.393} {\bibfield  {journal} {\bibinfo
  {journal} {Phys. Rev. Lett.}\ }\textbf {\bibinfo {volume} {85}},\ \bibinfo
  {pages} {393} (\bibinfo {year} {2000})}\BibitemShut {NoStop}%
\bibitem [{\citenamefont {Sinova}\ \emph {et~al.}(2015)\citenamefont {Sinova},
  \citenamefont {Valenzuela}, \citenamefont {Wunderlich}, \citenamefont
  {Back},\ and\ \citenamefont {Jungwirth}}]{RevModPhys.87.1213}%
  \BibitemOpen
  \bibfield  {author} {\bibinfo {author} {\bibfnamefont {J.}~\bibnamefont
  {Sinova}}, \bibinfo {author} {\bibfnamefont {S.~O.}\ \bibnamefont
  {Valenzuela}}, \bibinfo {author} {\bibfnamefont {J.}~\bibnamefont
  {Wunderlich}}, \bibinfo {author} {\bibfnamefont {C.~H.}\ \bibnamefont
  {Back}},\ and\ \bibinfo {author} {\bibfnamefont {T.}~\bibnamefont
  {Jungwirth}},\ }\bibfield  {title} {\bibinfo {title} {Spin {Hall} effects},\
  }\href {https://doi.org/10.1103/RevModPhys.87.1213} {\bibfield  {journal}
  {\bibinfo  {journal} {Rev. Mod. Phys.}\ }\textbf {\bibinfo {volume} {87}},\
  \bibinfo {pages} {1213} (\bibinfo {year} {2015})}\BibitemShut {NoStop}%
\bibitem [{\citenamefont {Liu}\ \emph {et~al.}(2011)\citenamefont {Liu},
  \citenamefont {Moriyama}, \citenamefont {Ralph},\ and\ \citenamefont
  {Buhrman}}]{PhysRevLett.106.036601}%
  \BibitemOpen
  \bibfield  {author} {\bibinfo {author} {\bibfnamefont {L.}~\bibnamefont
  {Liu}}, \bibinfo {author} {\bibfnamefont {T.}~\bibnamefont {Moriyama}},
  \bibinfo {author} {\bibfnamefont {D.~C.}\ \bibnamefont {Ralph}},\ and\
  \bibinfo {author} {\bibfnamefont {R.~A.}\ \bibnamefont {Buhrman}},\
  }\bibfield  {title} {\bibinfo {title} {Spin-torque ferromagnetic resonance
  induced by the spin {Hall} effect},\ }\href
  {https://doi.org/10.1103/PhysRevLett.106.036601} {\bibfield  {journal}
  {\bibinfo  {journal} {Phys. Rev. Lett.}\ }\textbf {\bibinfo {volume} {106}},\
  \bibinfo {pages} {036601} (\bibinfo {year} {2011})}\BibitemShut {NoStop}%
\bibitem [{\citenamefont {Liu}\ \emph {et~al.}(2012{\natexlab{a}})\citenamefont
  {Liu}, \citenamefont {Pai}, \citenamefont {Li}, \citenamefont {Tseng},
  \citenamefont {Ralph},\ and\ \citenamefont
  {Buhrman}}]{doi:10.1126/science.1218197}%
  \BibitemOpen
  \bibfield  {author} {\bibinfo {author} {\bibfnamefont {L.}~\bibnamefont
  {Liu}}, \bibinfo {author} {\bibfnamefont {C.-F.}\ \bibnamefont {Pai}},
  \bibinfo {author} {\bibfnamefont {Y.}~\bibnamefont {Li}}, \bibinfo {author}
  {\bibfnamefont {H.~W.}\ \bibnamefont {Tseng}}, \bibinfo {author}
  {\bibfnamefont {D.~C.}\ \bibnamefont {Ralph}},\ and\ \bibinfo {author}
  {\bibfnamefont {R.~A.}\ \bibnamefont {Buhrman}},\ }\bibfield  {title}
  {\bibinfo {title} {Spin-torque switching with the giant spin {Hall} effect of
  tantalum},\ }\href {https://doi.org/10.1126/science.1218197} {\bibfield
  {journal} {\bibinfo  {journal} {Science}\ }\textbf {\bibinfo {volume}
  {336}},\ \bibinfo {pages} {555} (\bibinfo {year}
  {2012}{\natexlab{a}})}\BibitemShut {NoStop}%
\bibitem [{\citenamefont {Liu}\ \emph {et~al.}(2012{\natexlab{b}})\citenamefont
  {Liu}, \citenamefont {Lee}, \citenamefont {Gudmundsen}, \citenamefont
  {Ralph},\ and\ \citenamefont {Buhrman}}]{PhysRevLett.109.096602}%
  \BibitemOpen
  \bibfield  {author} {\bibinfo {author} {\bibfnamefont {L.}~\bibnamefont
  {Liu}}, \bibinfo {author} {\bibfnamefont {O.~J.}\ \bibnamefont {Lee}},
  \bibinfo {author} {\bibfnamefont {T.~J.}\ \bibnamefont {Gudmundsen}},
  \bibinfo {author} {\bibfnamefont {D.~C.}\ \bibnamefont {Ralph}},\ and\
  \bibinfo {author} {\bibfnamefont {R.~A.}\ \bibnamefont {Buhrman}},\
  }\bibfield  {title} {\bibinfo {title} {Current-induced switching of
  perpendicularly magnetized magnetic layers using spin torque from the spin
  {Hall} effect},\ }\href {https://doi.org/10.1103/PhysRevLett.109.096602}
  {\bibfield  {journal} {\bibinfo  {journal} {Phys. Rev. Lett.}\ }\textbf
  {\bibinfo {volume} {109}},\ \bibinfo {pages} {096602} (\bibinfo {year}
  {2012}{\natexlab{b}})}\BibitemShut {NoStop}%
\bibitem [{\citenamefont {Hoffmann}(2013)}]{6516040}%
  \BibitemOpen
  \bibfield  {author} {\bibinfo {author} {\bibfnamefont {A.}~\bibnamefont
  {Hoffmann}},\ }\bibfield  {title} {\bibinfo {title} {Spin {Hall} effects in
  metals},\ }\href {https://doi.org/10.1109/TMAG.2013.2262947} {\bibfield
  {journal} {\bibinfo  {journal} {IEEE Trans. Magn,}\ }\textbf {\bibinfo
  {volume} {49}},\ \bibinfo {pages} {5172} (\bibinfo {year}
  {2013})}\BibitemShut {NoStop}%
\bibitem [{\citenamefont {Zhang}\ \emph {et~al.}(2016)\citenamefont {Zhang},
  \citenamefont {Sklenar}, \citenamefont {Hsu}, \citenamefont {Jiang},
  \citenamefont {Jungfleisch}, \citenamefont {Xiao}, \citenamefont {Fradin},
  \citenamefont {Liu}, \citenamefont {Pearson}, \citenamefont {Ketterson},
  \citenamefont {Yang},\ and\ \citenamefont {Hoffmann}}]{zhang_research_2016}%
  \BibitemOpen
  \bibfield  {author} {\bibinfo {author} {\bibfnamefont {W.}~\bibnamefont
  {Zhang}}, \bibinfo {author} {\bibfnamefont {J.}~\bibnamefont {Sklenar}},
  \bibinfo {author} {\bibfnamefont {B.}~\bibnamefont {Hsu}}, \bibinfo {author}
  {\bibfnamefont {W.}~\bibnamefont {Jiang}}, \bibinfo {author} {\bibfnamefont
  {M.~B.}\ \bibnamefont {Jungfleisch}}, \bibinfo {author} {\bibfnamefont
  {J.}~\bibnamefont {Xiao}}, \bibinfo {author} {\bibfnamefont {F.~Y.}\
  \bibnamefont {Fradin}}, \bibinfo {author} {\bibfnamefont {Y.}~\bibnamefont
  {Liu}}, \bibinfo {author} {\bibfnamefont {J.~E.}\ \bibnamefont {Pearson}},
  \bibinfo {author} {\bibfnamefont {J.~B.}\ \bibnamefont {Ketterson}}, \bibinfo
  {author} {\bibfnamefont {Z.}~\bibnamefont {Yang}},\ and\ \bibinfo {author}
  {\bibfnamefont {A.}~\bibnamefont {Hoffmann}},\ }\bibfield  {title} {\bibinfo
  {title} {Research {Update}: {Spin} transfer torques in permalloy on monolayer
  {MoS$_2$}},\ }\href@noop {} {\bibfield  {journal} {\bibinfo  {journal} {APL
  Mater.}\ }\textbf {\bibinfo {volume} {4}},\ \bibinfo {pages} {032302}
  (\bibinfo {year} {2016})}\BibitemShut {NoStop}%
\bibitem [{\citenamefont {Shao}\ \emph {et~al.}(2016)\citenamefont {Shao},
  \citenamefont {Yu}, \citenamefont {Lan}, \citenamefont {Shi}, \citenamefont
  {Li}, \citenamefont {Zheng}, \citenamefont {Zhu}, \citenamefont {Li},
  \citenamefont {Amiri},\ and\ \citenamefont
  {Wang}}]{doi:10.1021/acs.nanolett.6b03300}%
  \BibitemOpen
  \bibfield  {author} {\bibinfo {author} {\bibfnamefont {Q.}~\bibnamefont
  {Shao}}, \bibinfo {author} {\bibfnamefont {G.}~\bibnamefont {Yu}}, \bibinfo
  {author} {\bibfnamefont {Y.-W.}\ \bibnamefont {Lan}}, \bibinfo {author}
  {\bibfnamefont {Y.}~\bibnamefont {Shi}}, \bibinfo {author} {\bibfnamefont
  {M.-Y.}\ \bibnamefont {Li}}, \bibinfo {author} {\bibfnamefont
  {C.}~\bibnamefont {Zheng}}, \bibinfo {author} {\bibfnamefont
  {X.}~\bibnamefont {Zhu}}, \bibinfo {author} {\bibfnamefont {L.-J.}\
  \bibnamefont {Li}}, \bibinfo {author} {\bibfnamefont {P.~K.}\ \bibnamefont
  {Amiri}},\ and\ \bibinfo {author} {\bibfnamefont {K.~L.}\ \bibnamefont
  {Wang}},\ }\bibfield  {title} {\bibinfo {title} {Strong {Rashba-Edelstein}
  effect-induced spin–orbit torques in monolayer transition metal
  dichalcogenide/ferromagnet bilayers},\ }\href
  {https://doi.org/10.1021/acs.nanolett.6b03300} {\bibfield  {journal}
  {\bibinfo  {journal} {Nano Lett.}\ }\textbf {\bibinfo {volume} {16}},\
  \bibinfo {pages} {7514} (\bibinfo {year} {2016})}\BibitemShut {NoStop}%
\bibitem [{\citenamefont {Emori}\ \emph {et~al.}(2016)\citenamefont {Emori},
  \citenamefont {Nan}, \citenamefont {Belkessam}, \citenamefont {Wang},
  \citenamefont {Matyushov}, \citenamefont {Babroski}, \citenamefont {Gao},
  \citenamefont {Lin},\ and\ \citenamefont {Sun}}]{PhysRevB.93.180402}%
  \BibitemOpen
  \bibfield  {author} {\bibinfo {author} {\bibfnamefont {S.}~\bibnamefont
  {Emori}}, \bibinfo {author} {\bibfnamefont {T.}~\bibnamefont {Nan}}, \bibinfo
  {author} {\bibfnamefont {A.~M.}\ \bibnamefont {Belkessam}}, \bibinfo {author}
  {\bibfnamefont {X.}~\bibnamefont {Wang}}, \bibinfo {author} {\bibfnamefont
  {A.~D.}\ \bibnamefont {Matyushov}}, \bibinfo {author} {\bibfnamefont {C.~J.}\
  \bibnamefont {Babroski}}, \bibinfo {author} {\bibfnamefont {Y.}~\bibnamefont
  {Gao}}, \bibinfo {author} {\bibfnamefont {H.}~\bibnamefont {Lin}},\ and\
  \bibinfo {author} {\bibfnamefont {N.~X.}\ \bibnamefont {Sun}},\ }\bibfield
  {title} {\bibinfo {title} {Interfacial spin-orbit torque without bulk
  spin-orbit coupling},\ }\href {https://doi.org/10.1103/PhysRevB.93.180402}
  {\bibfield  {journal} {\bibinfo  {journal} {Phys. Rev. B}\ }\textbf {\bibinfo
  {volume} {93}},\ \bibinfo {pages} {180402} (\bibinfo {year}
  {2016})}\BibitemShut {NoStop}%
\bibitem [{\citenamefont {Mellnik}\ \emph {et~al.}(2014)\citenamefont
  {Mellnik}, \citenamefont {Lee}, \citenamefont {Richardella}, \citenamefont
  {Grab}, \citenamefont {Mintun}, \citenamefont {Fischer}, \citenamefont
  {Vaezi}, \citenamefont {Manchon}, \citenamefont {Kim}, \citenamefont
  {Samarth},\ and\ \citenamefont {Ralph}}]{mellnik_spin-transfer_2014}%
  \BibitemOpen
  \bibfield  {author} {\bibinfo {author} {\bibfnamefont {A.~R.}\ \bibnamefont
  {Mellnik}}, \bibinfo {author} {\bibfnamefont {J.~S.}\ \bibnamefont {Lee}},
  \bibinfo {author} {\bibfnamefont {A.}~\bibnamefont {Richardella}}, \bibinfo
  {author} {\bibfnamefont {J.~L.}\ \bibnamefont {Grab}}, \bibinfo {author}
  {\bibfnamefont {P.~J.}\ \bibnamefont {Mintun}}, \bibinfo {author}
  {\bibfnamefont {M.~H.}\ \bibnamefont {Fischer}}, \bibinfo {author}
  {\bibfnamefont {A.}~\bibnamefont {Vaezi}}, \bibinfo {author} {\bibfnamefont
  {A.}~\bibnamefont {Manchon}}, \bibinfo {author} {\bibfnamefont {E.-A.}\
  \bibnamefont {Kim}}, \bibinfo {author} {\bibfnamefont {N.}~\bibnamefont
  {Samarth}},\ and\ \bibinfo {author} {\bibfnamefont {D.~C.}\ \bibnamefont
  {Ralph}},\ }\bibfield  {title} {\bibinfo {title} {Spin-transfer torque
  generated by a topological insulator},\ }\href
  {https://doi.org/10.1038/nature13534} {\bibfield  {journal} {\bibinfo
  {journal} {Nature}\ }\textbf {\bibinfo {volume} {511}},\ \bibinfo {pages}
  {449} (\bibinfo {year} {2014})}\BibitemShut {NoStop}%
\bibitem [{\citenamefont {Mihai~Miron}\ \emph {et~al.}(2010)\citenamefont
  {Mihai~Miron}, \citenamefont {Gaudin}, \citenamefont {Auffret}, \citenamefont
  {Rodmacq}, \citenamefont {Schuhl}, \citenamefont {Pizzini}, \citenamefont
  {Vogel},\ and\ \citenamefont
  {Gambardella}}]{mihai_miron_current-driven_2010}%
  \BibitemOpen
  \bibfield  {author} {\bibinfo {author} {\bibfnamefont {I.}~\bibnamefont
  {Mihai~Miron}}, \bibinfo {author} {\bibfnamefont {G.}~\bibnamefont {Gaudin}},
  \bibinfo {author} {\bibfnamefont {S.}~\bibnamefont {Auffret}}, \bibinfo
  {author} {\bibfnamefont {B.}~\bibnamefont {Rodmacq}}, \bibinfo {author}
  {\bibfnamefont {A.}~\bibnamefont {Schuhl}}, \bibinfo {author} {\bibfnamefont
  {S.}~\bibnamefont {Pizzini}}, \bibinfo {author} {\bibfnamefont
  {J.}~\bibnamefont {Vogel}},\ and\ \bibinfo {author} {\bibfnamefont
  {P.}~\bibnamefont {Gambardella}},\ }\bibfield  {title} {\bibinfo {title}
  {Current-driven spin torque induced by the {Rashba} effect in a ferromagnetic
  metal layer},\ }\href {https://doi.org/10.1038/nmat2613} {\bibfield
  {journal} {\bibinfo  {journal} {Nat. Mater.}\ }\textbf {\bibinfo {volume}
  {9}},\ \bibinfo {pages} {230} (\bibinfo {year} {2010})}\BibitemShut {NoStop}%
\bibitem [{\citenamefont {MacNeill}\ \emph {et~al.}(2017)\citenamefont
  {MacNeill}, \citenamefont {Stiehl}, \citenamefont {Guimaraes}, \citenamefont
  {Buhrman}, \citenamefont {Park},\ and\ \citenamefont
  {Ralph}}]{macneill_control_2017}%
  \BibitemOpen
  \bibfield  {author} {\bibinfo {author} {\bibfnamefont {D.}~\bibnamefont
  {MacNeill}}, \bibinfo {author} {\bibfnamefont {G.~M.}\ \bibnamefont
  {Stiehl}}, \bibinfo {author} {\bibfnamefont {M.~H.~D.}\ \bibnamefont
  {Guimaraes}}, \bibinfo {author} {\bibfnamefont {R.~A.}\ \bibnamefont
  {Buhrman}}, \bibinfo {author} {\bibfnamefont {J.}~\bibnamefont {Park}},\ and\
  \bibinfo {author} {\bibfnamefont {D.~C.}\ \bibnamefont {Ralph}},\ }\bibfield
  {title} {\bibinfo {title} {Control of spin–orbit torques through crystal
  symmetry in {WTe$_2$}/ferromagnet bilayers},\ }\href
  {https://doi.org/10.1038/nphys3933} {\bibfield  {journal} {\bibinfo
  {journal} {Nat. Phys.}\ }\textbf {\bibinfo {volume} {13}},\ \bibinfo {pages}
  {300} (\bibinfo {year} {2017})}\BibitemShut {NoStop}%
\bibitem [{\citenamefont {Li}\ \emph {et~al.}(2018)\citenamefont {Li},
  \citenamefont {Wu}, \citenamefont {Wen}, \citenamefont {Zhang}, \citenamefont
  {Zhang}, \citenamefont {Zhang}, \citenamefont {Yu}, \citenamefont {Yang},
  \citenamefont {Manchon},\ and\ \citenamefont
  {Zhang}}]{li_spin-momentum_2018}%
  \BibitemOpen
  \bibfield  {author} {\bibinfo {author} {\bibfnamefont {P.}~\bibnamefont
  {Li}}, \bibinfo {author} {\bibfnamefont {W.}~\bibnamefont {Wu}}, \bibinfo
  {author} {\bibfnamefont {Y.}~\bibnamefont {Wen}}, \bibinfo {author}
  {\bibfnamefont {C.}~\bibnamefont {Zhang}}, \bibinfo {author} {\bibfnamefont
  {J.}~\bibnamefont {Zhang}}, \bibinfo {author} {\bibfnamefont
  {S.}~\bibnamefont {Zhang}}, \bibinfo {author} {\bibfnamefont
  {Z.}~\bibnamefont {Yu}}, \bibinfo {author} {\bibfnamefont {S.~A.}\
  \bibnamefont {Yang}}, \bibinfo {author} {\bibfnamefont {A.}~\bibnamefont
  {Manchon}},\ and\ \bibinfo {author} {\bibfnamefont {X.-x.}\ \bibnamefont
  {Zhang}},\ }\bibfield  {title} {\bibinfo {title} {Spin-momentum locking and
  spin-orbit torques in magnetic nano-heterojunctions composed of {Weyl}
  semimetal {WTe$_2$}},\ }\href {https://doi.org/10.1038/s41467-018-06518-1}
  {\bibfield  {journal} {\bibinfo  {journal} {Nat. Commun.}\ }\textbf {\bibinfo
  {volume} {9}},\ \bibinfo {pages} {3990} (\bibinfo {year} {2018})}\BibitemShut
  {NoStop}%
\bibitem [{\citenamefont {Dc}\ \emph {et~al.}(2023)\citenamefont {Dc},
  \citenamefont {Shao}, \citenamefont {Hou}, \citenamefont {Vailionis},
  \citenamefont {Quarterman}, \citenamefont {Habiboglu}, \citenamefont
  {Venuti}, \citenamefont {Xue}, \citenamefont {Huang}, \citenamefont {Lee},
  \citenamefont {Miura}, \citenamefont {Kirby}, \citenamefont {Bi},
  \citenamefont {Li}, \citenamefont {Deng}, \citenamefont {Lin}, \citenamefont
  {Tsai}, \citenamefont {Eley}, \citenamefont {Wang}, \citenamefont {Borchers},
  \citenamefont {Tsymbal},\ and\ \citenamefont {Wang}}]{dc_observation_2023}%
  \BibitemOpen
  \bibfield  {author} {\bibinfo {author} {\bibfnamefont {M.}~\bibnamefont
  {Dc}}, \bibinfo {author} {\bibfnamefont {D.-F.}\ \bibnamefont {Shao}},
  \bibinfo {author} {\bibfnamefont {V.~D.-H.}\ \bibnamefont {Hou}}, \bibinfo
  {author} {\bibfnamefont {A.}~\bibnamefont {Vailionis}}, \bibinfo {author}
  {\bibfnamefont {P.}~\bibnamefont {Quarterman}}, \bibinfo {author}
  {\bibfnamefont {A.}~\bibnamefont {Habiboglu}}, \bibinfo {author}
  {\bibfnamefont {M.~B.}\ \bibnamefont {Venuti}}, \bibinfo {author}
  {\bibfnamefont {F.}~\bibnamefont {Xue}}, \bibinfo {author} {\bibfnamefont
  {Y.-L.}\ \bibnamefont {Huang}}, \bibinfo {author} {\bibfnamefont {C.-M.}\
  \bibnamefont {Lee}}, \bibinfo {author} {\bibfnamefont {M.}~\bibnamefont
  {Miura}}, \bibinfo {author} {\bibfnamefont {B.}~\bibnamefont {Kirby}},
  \bibinfo {author} {\bibfnamefont {C.}~\bibnamefont {Bi}}, \bibinfo {author}
  {\bibfnamefont {X.}~\bibnamefont {Li}}, \bibinfo {author} {\bibfnamefont
  {Y.}~\bibnamefont {Deng}}, \bibinfo {author} {\bibfnamefont {S.-J.}\
  \bibnamefont {Lin}}, \bibinfo {author} {\bibfnamefont {W.}~\bibnamefont
  {Tsai}}, \bibinfo {author} {\bibfnamefont {S.}~\bibnamefont {Eley}}, \bibinfo
  {author} {\bibfnamefont {W.-G.}\ \bibnamefont {Wang}}, \bibinfo {author}
  {\bibfnamefont {J.~A.}\ \bibnamefont {Borchers}}, \bibinfo {author}
  {\bibfnamefont {E.~Y.}\ \bibnamefont {Tsymbal}},\ and\ \bibinfo {author}
  {\bibfnamefont {S.~X.}\ \bibnamefont {Wang}},\ }\bibfield  {title} {\bibinfo
  {title} {Observation of anti-damping spin–orbit torques generated by
  in-plane and out-of-plane spin polarizations in {MnPd$_3$}},\ }\href
  {https://doi.org/10.1038/s41563-023-01522-3} {\bibfield  {journal} {\bibinfo
  {journal} {Nat. Mater.}\ ,\ \bibinfo {pages} {1}} (\bibinfo {year}
  {2023})}\BibitemShut {NoStop}%
\bibitem [{\citenamefont {Guimarães}\ \emph {et~al.}(2018)\citenamefont
  {Guimarães}, \citenamefont {Stiehl}, \citenamefont {MacNeill}, \citenamefont
  {Reynolds},\ and\ \citenamefont {Ralph}}]{doi:10.1021/acs.nanolett.7b04993}%
  \BibitemOpen
  \bibfield  {author} {\bibinfo {author} {\bibfnamefont {M.~H.~D.}\
  \bibnamefont {Guimarães}}, \bibinfo {author} {\bibfnamefont {G.~M.}\
  \bibnamefont {Stiehl}}, \bibinfo {author} {\bibfnamefont {D.}~\bibnamefont
  {MacNeill}}, \bibinfo {author} {\bibfnamefont {N.~D.}\ \bibnamefont
  {Reynolds}},\ and\ \bibinfo {author} {\bibfnamefont {D.~C.}\ \bibnamefont
  {Ralph}},\ }\bibfield  {title} {\bibinfo {title} {Spin–orbit torques in
  {NbSe$_2$}/permalloy bilayers},\ }\href
  {https://doi.org/10.1021/acs.nanolett.7b04993} {\bibfield  {journal}
  {\bibinfo  {journal} {Nano Lett.}\ }\textbf {\bibinfo {volume} {18}},\
  \bibinfo {pages} {1311} (\bibinfo {year} {2018})}\BibitemShut {NoStop}%
\bibitem [{\citenamefont {Avci}\ \emph {et~al.}(2014)\citenamefont {Avci},
  \citenamefont {Garello}, \citenamefont {Gabureac}, \citenamefont {Ghosh},
  \citenamefont {Fuhrer}, \citenamefont {Alvarado},\ and\ \citenamefont
  {Gambardella}}]{PhysRevB.90.224427}%
  \BibitemOpen
  \bibfield  {author} {\bibinfo {author} {\bibfnamefont {C.~O.}\ \bibnamefont
  {Avci}}, \bibinfo {author} {\bibfnamefont {K.}~\bibnamefont {Garello}},
  \bibinfo {author} {\bibfnamefont {M.}~\bibnamefont {Gabureac}}, \bibinfo
  {author} {\bibfnamefont {A.}~\bibnamefont {Ghosh}}, \bibinfo {author}
  {\bibfnamefont {A.}~\bibnamefont {Fuhrer}}, \bibinfo {author} {\bibfnamefont
  {S.~F.}\ \bibnamefont {Alvarado}},\ and\ \bibinfo {author} {\bibfnamefont
  {P.}~\bibnamefont {Gambardella}},\ }\bibfield  {title} {\bibinfo {title}
  {Interplay of spin-orbit torque and thermoelectric effects in
  ferromagnet/normal-metal bilayers},\ }\href
  {https://doi.org/10.1103/PhysRevB.90.224427} {\bibfield  {journal} {\bibinfo
  {journal} {Phys. Rev. B}\ }\textbf {\bibinfo {volume} {90}},\ \bibinfo
  {pages} {224427} (\bibinfo {year} {2014})}\BibitemShut {NoStop}%
\bibitem [{\citenamefont {Hayashi}\ \emph {et~al.}(2014)\citenamefont
  {Hayashi}, \citenamefont {Kim}, \citenamefont {Yamanouchi},\ and\
  \citenamefont {Ohno}}]{PhysRevB.89.144425}%
  \BibitemOpen
  \bibfield  {author} {\bibinfo {author} {\bibfnamefont {M.}~\bibnamefont
  {Hayashi}}, \bibinfo {author} {\bibfnamefont {J.}~\bibnamefont {Kim}},
  \bibinfo {author} {\bibfnamefont {M.}~\bibnamefont {Yamanouchi}},\ and\
  \bibinfo {author} {\bibfnamefont {H.}~\bibnamefont {Ohno}},\ }\bibfield
  {title} {\bibinfo {title} {Quantitative characterization of the spin-orbit
  torque using harmonic {Hall} voltage measurements},\ }\href
  {https://doi.org/10.1103/PhysRevB.89.144425} {\bibfield  {journal} {\bibinfo
  {journal} {Phys. Rev. B}\ }\textbf {\bibinfo {volume} {89}},\ \bibinfo
  {pages} {144425} (\bibinfo {year} {2014})}\BibitemShut {NoStop}%
\bibitem [{\citenamefont {Pace}\ \emph {et~al.}(2021)\citenamefont {Pace},
  \citenamefont {Martini}, \citenamefont {Convertino}, \citenamefont {Keum},
  \citenamefont {Forti}, \citenamefont {Pezzini}, \citenamefont {Fabbri},
  \citenamefont {Mišeikis},\ and\ \citenamefont
  {Coletti}}]{doi:10.1021/acsnano.0c05936}%
  \BibitemOpen
  \bibfield  {author} {\bibinfo {author} {\bibfnamefont {S.}~\bibnamefont
  {Pace}}, \bibinfo {author} {\bibfnamefont {L.}~\bibnamefont {Martini}},
  \bibinfo {author} {\bibfnamefont {D.}~\bibnamefont {Convertino}}, \bibinfo
  {author} {\bibfnamefont {D.~H.}\ \bibnamefont {Keum}}, \bibinfo {author}
  {\bibfnamefont {S.}~\bibnamefont {Forti}}, \bibinfo {author} {\bibfnamefont
  {S.}~\bibnamefont {Pezzini}}, \bibinfo {author} {\bibfnamefont
  {F.}~\bibnamefont {Fabbri}}, \bibinfo {author} {\bibfnamefont
  {V.}~\bibnamefont {Mišeikis}},\ and\ \bibinfo {author} {\bibfnamefont
  {C.}~\bibnamefont {Coletti}},\ }\bibfield  {title} {\bibinfo {title}
  {Synthesis of large-scale monolayer {1T'-MoTe$_2$} and its stabilization via
  scalable {hBN} encapsulation},\ }\href
  {https://doi.org/10.1021/acsnano.0c05936} {\bibfield  {journal} {\bibinfo
  {journal} {ACS Nano}\ }\textbf {\bibinfo {volume} {15}},\ \bibinfo {pages}
  {4213} (\bibinfo {year} {2021})}\BibitemShut {NoStop}%
\bibitem [{\citenamefont {Karki}\ \emph {et~al.}(2020)\citenamefont {Karki},
  \citenamefont {Freelon}, \citenamefont {Rajapakse}, \citenamefont {Musa},
  \citenamefont {Riyadh}, \citenamefont {Morris}, \citenamefont {Abu},
  \citenamefont {Yu}, \citenamefont {Sumanasekera},\ and\ \citenamefont
  {Jasinski}}]{karki_strain-induced_2020}%
  \BibitemOpen
  \bibfield  {author} {\bibinfo {author} {\bibfnamefont {B.}~\bibnamefont
  {Karki}}, \bibinfo {author} {\bibfnamefont {B.}~\bibnamefont {Freelon}},
  \bibinfo {author} {\bibfnamefont {M.}~\bibnamefont {Rajapakse}}, \bibinfo
  {author} {\bibfnamefont {R.}~\bibnamefont {Musa}}, \bibinfo {author}
  {\bibfnamefont {S.~M.~S.}\ \bibnamefont {Riyadh}}, \bibinfo {author}
  {\bibfnamefont {B.}~\bibnamefont {Morris}}, \bibinfo {author} {\bibfnamefont
  {U.}~\bibnamefont {Abu}}, \bibinfo {author} {\bibfnamefont {M.}~\bibnamefont
  {Yu}}, \bibinfo {author} {\bibfnamefont {G.}~\bibnamefont {Sumanasekera}},\
  and\ \bibinfo {author} {\bibfnamefont {J.~B.}\ \bibnamefont {Jasinski}},\
  }\bibfield  {title} {\bibinfo {title} {Strain-induced vibrational properties
  of few layer black phosphorus and {MoTe$_2$} via {Raman} spectroscopy},\
  }\href {https://doi.org/10.1088/1361-6528/aba13e} {\bibfield  {journal}
  {\bibinfo  {journal} {Nanotechn.}\ }\textbf {\bibinfo {volume} {31}},\
  \bibinfo {pages} {425707} (\bibinfo {year} {2020})}\BibitemShut {NoStop}%
\bibitem [{\citenamefont {Imajo}\ \emph {et~al.}(2021)\citenamefont {Imajo},
  \citenamefont {Suemasu},\ and\ \citenamefont {Toko}}]{imajo_strain_2021}%
  \BibitemOpen
  \bibfield  {author} {\bibinfo {author} {\bibfnamefont {T.}~\bibnamefont
  {Imajo}}, \bibinfo {author} {\bibfnamefont {T.}~\bibnamefont {Suemasu}},\
  and\ \bibinfo {author} {\bibfnamefont {K.}~\bibnamefont {Toko}},\ }\bibfield
  {title} {\bibinfo {title} {Strain effects on polycrystalline germanium thin
  films},\ }\href {https://doi.org/10.1038/s41598-021-87616-x} {\bibfield
  {journal} {\bibinfo  {journal} {Sci. Rep.}\ }\textbf {\bibinfo {volume}
  {11}},\ \bibinfo {pages} {8333} (\bibinfo {year} {2021})}\BibitemShut
  {NoStop}%
\bibitem [{\citenamefont {Angel}\ \emph {et~al.}(2019)\citenamefont {Angel},
  \citenamefont {Murri}, \citenamefont {Mihailova},\ and\ \citenamefont
  {Alvaro}}]{AngelMurriMihailovaAlvaro+2019+129+140}%
  \BibitemOpen
  \bibfield  {author} {\bibinfo {author} {\bibfnamefont {R.~J.}\ \bibnamefont
  {Angel}}, \bibinfo {author} {\bibfnamefont {M.}~\bibnamefont {Murri}},
  \bibinfo {author} {\bibfnamefont {B.}~\bibnamefont {Mihailova}},\ and\
  \bibinfo {author} {\bibfnamefont {M.}~\bibnamefont {Alvaro}},\ }\bibfield
  {title} {\bibinfo {title} {Stress, strain and {Raman} shifts},\ }\href
  {https://doi.org/doi:10.1515/zkri-2018-2112} {\bibfield  {journal} {\bibinfo
  {journal} {Z. Kristallogr. Cryst. Mater.}\ }\textbf {\bibinfo {volume}
  {234}},\ \bibinfo {pages} {129} (\bibinfo {year} {2019})}\BibitemShut
  {NoStop}%
\bibitem [{\citenamefont {Ma}\ \emph {et~al.}(2016)\citenamefont {Ma},
  \citenamefont {Guo}, \citenamefont {Yi}, \citenamefont {Yu}, \citenamefont
  {Zhang}, \citenamefont {Ji}, \citenamefont {Tian}, \citenamefont {Jin},
  \citenamefont {Wang}, \citenamefont {Liu}, \citenamefont {Xia}, \citenamefont
  {Shi},\ and\ \citenamefont {Zhang}}]{PhysRevB.94.214105}%
  \BibitemOpen
  \bibfield  {author} {\bibinfo {author} {\bibfnamefont {X.}~\bibnamefont
  {Ma}}, \bibinfo {author} {\bibfnamefont {P.}~\bibnamefont {Guo}}, \bibinfo
  {author} {\bibfnamefont {C.}~\bibnamefont {Yi}}, \bibinfo {author}
  {\bibfnamefont {Q.}~\bibnamefont {Yu}}, \bibinfo {author} {\bibfnamefont
  {A.}~\bibnamefont {Zhang}}, \bibinfo {author} {\bibfnamefont
  {J.}~\bibnamefont {Ji}}, \bibinfo {author} {\bibfnamefont {Y.}~\bibnamefont
  {Tian}}, \bibinfo {author} {\bibfnamefont {F.}~\bibnamefont {Jin}}, \bibinfo
  {author} {\bibfnamefont {Y.}~\bibnamefont {Wang}}, \bibinfo {author}
  {\bibfnamefont {K.}~\bibnamefont {Liu}}, \bibinfo {author} {\bibfnamefont
  {T.}~\bibnamefont {Xia}}, \bibinfo {author} {\bibfnamefont {Y.}~\bibnamefont
  {Shi}},\ and\ \bibinfo {author} {\bibfnamefont {Q.}~\bibnamefont {Zhang}},\
  }\bibfield  {title} {\bibinfo {title} {Raman scattering in the
  transition-metal dichalcogenides of
  $1{T}^{\ensuremath{'}}\text{\ensuremath{-}}\mathrm{MoT}{\mathrm{e}}_{2},{T}_{d}\text{\ensuremath{-}}\mathrm{MoT}{\mathrm{e}}_{2}$,
  and ${T}_{d}\text{\ensuremath{-}}\mathrm{WT}{\mathrm{e}}_{2}$},\ }\href
  {https://doi.org/10.1103/PhysRevB.94.214105} {\bibfield  {journal} {\bibinfo
  {journal} {Phys. Rev. B}\ }\textbf {\bibinfo {volume} {94}},\ \bibinfo
  {pages} {214105} (\bibinfo {year} {2016})}\BibitemShut {NoStop}%
\bibitem [{\citenamefont {Beams}\ \emph {et~al.}(2016)\citenamefont {Beams},
  \citenamefont {Cançado}, \citenamefont {Krylyuk}, \citenamefont {Kalish},
  \citenamefont {Kalanyan}, \citenamefont {Singh}, \citenamefont {Choudhary},
  \citenamefont {Bruma}, \citenamefont {Vora}, \citenamefont {Tavazza},
  \citenamefont {Davydov},\ and\ \citenamefont
  {Stranick}}]{doi:10.1021/acsnano.6b05127}%
  \BibitemOpen
  \bibfield  {author} {\bibinfo {author} {\bibfnamefont {R.}~\bibnamefont
  {Beams}}, \bibinfo {author} {\bibfnamefont {L.~G.}\ \bibnamefont {Cançado}},
  \bibinfo {author} {\bibfnamefont {S.}~\bibnamefont {Krylyuk}}, \bibinfo
  {author} {\bibfnamefont {I.}~\bibnamefont {Kalish}}, \bibinfo {author}
  {\bibfnamefont {B.}~\bibnamefont {Kalanyan}}, \bibinfo {author}
  {\bibfnamefont {A.~K.}\ \bibnamefont {Singh}}, \bibinfo {author}
  {\bibfnamefont {K.}~\bibnamefont {Choudhary}}, \bibinfo {author}
  {\bibfnamefont {A.}~\bibnamefont {Bruma}}, \bibinfo {author} {\bibfnamefont
  {P.~M.}\ \bibnamefont {Vora}}, \bibinfo {author} {\bibfnamefont
  {F.}~\bibnamefont {Tavazza}}, \bibinfo {author} {\bibfnamefont {A.~V.}\
  \bibnamefont {Davydov}},\ and\ \bibinfo {author} {\bibfnamefont {S.~J.}\
  \bibnamefont {Stranick}},\ }\bibfield  {title} {\bibinfo {title}
  {Characterization of few-layer {1T' MoTe$_2$} by polarization-resolved second
  harmonic generation and {Raman} scattering},\ }\href
  {https://doi.org/10.1021/acsnano.6b05127} {\bibfield  {journal} {\bibinfo
  {journal} {ACS Nano}\ }\textbf {\bibinfo {volume} {10}},\ \bibinfo {pages}
  {9626} (\bibinfo {year} {2016})}\BibitemShut {NoStop}%
\bibitem [{\citenamefont {Mosendz}\ \emph {et~al.}(2010)\citenamefont
  {Mosendz}, \citenamefont {Pearson}, \citenamefont {Fradin}, \citenamefont
  {Bauer}, \citenamefont {Bader},\ and\ \citenamefont
  {Hoffmann}}]{PhysRevLett.104.046601}%
  \BibitemOpen
  \bibfield  {author} {\bibinfo {author} {\bibfnamefont {O.}~\bibnamefont
  {Mosendz}}, \bibinfo {author} {\bibfnamefont {J.~E.}\ \bibnamefont
  {Pearson}}, \bibinfo {author} {\bibfnamefont {F.~Y.}\ \bibnamefont {Fradin}},
  \bibinfo {author} {\bibfnamefont {G.~E.~W.}\ \bibnamefont {Bauer}}, \bibinfo
  {author} {\bibfnamefont {S.~D.}\ \bibnamefont {Bader}},\ and\ \bibinfo
  {author} {\bibfnamefont {A.}~\bibnamefont {Hoffmann}},\ }\bibfield  {title}
  {\bibinfo {title} {Quantifying spin hall angles from spin pumping:
  Experiments and theory},\ }\href
  {https://doi.org/10.1103/PhysRevLett.104.046601} {\bibfield  {journal}
  {\bibinfo  {journal} {Phys. Rev. Lett.}\ }\textbf {\bibinfo {volume} {104}},\
  \bibinfo {pages} {046601} (\bibinfo {year} {2010})}\BibitemShut {NoStop}%
\bibitem [{\citenamefont {Tian}\ \emph {et~al.}(2016)\citenamefont {Tian},
  \citenamefont {Yang}, \citenamefont {Liu}, \citenamefont {Wang},
  \citenamefont {Pan}, \citenamefont {Gu},\ and\ \citenamefont
  {Li}}]{tian_temperature-dependent_2016}%
  \BibitemOpen
  \bibfield  {author} {\bibinfo {author} {\bibfnamefont {S.}~\bibnamefont
  {Tian}}, \bibinfo {author} {\bibfnamefont {Y.}~\bibnamefont {Yang}}, \bibinfo
  {author} {\bibfnamefont {Z.}~\bibnamefont {Liu}}, \bibinfo {author}
  {\bibfnamefont {C.}~\bibnamefont {Wang}}, \bibinfo {author} {\bibfnamefont
  {R.}~\bibnamefont {Pan}}, \bibinfo {author} {\bibfnamefont {C.}~\bibnamefont
  {Gu}},\ and\ \bibinfo {author} {\bibfnamefont {J.}~\bibnamefont {Li}},\
  }\bibfield  {title} {\bibinfo {title} {Temperature-dependent {Raman}
  investigation on suspended graphene: {Contribution} from thermal expansion
  coefficient mismatch between graphene and substrate},\ }\href
  {https://doi.org/https://doi.org/10.1016/j.carbon.2016.03.046} {\bibfield
  {journal} {\bibinfo  {journal} {Carbon}\ }\textbf {\bibinfo {volume} {104}},\
  \bibinfo {pages} {27} (\bibinfo {year} {2016})}\BibitemShut {NoStop}%
\bibitem [{\citenamefont {Yoon}\ \emph {et~al.}(2011)\citenamefont {Yoon},
  \citenamefont {Son},\ and\ \citenamefont {Cheong}}]{doi:10.1021/nl201488g}%
  \BibitemOpen
  \bibfield  {author} {\bibinfo {author} {\bibfnamefont {D.}~\bibnamefont
  {Yoon}}, \bibinfo {author} {\bibfnamefont {Y.-W.}\ \bibnamefont {Son}},\ and\
  \bibinfo {author} {\bibfnamefont {H.}~\bibnamefont {Cheong}},\ }\bibfield
  {title} {\bibinfo {title} {Negative thermal expansion coefficient of graphene
  measured by {Raman} spectroscopy},\ }\href
  {https://doi.org/10.1021/nl201488g} {\bibfield  {journal} {\bibinfo
  {journal} {Nano Lett.}\ }\textbf {\bibinfo {volume} {11}},\ \bibinfo {pages}
  {3227} (\bibinfo {year} {2011})}\BibitemShut {NoStop}%
\bibitem [{\citenamefont {Chyczewski}\ \emph {et~al.}(2024)\citenamefont
  {Chyczewski}, \citenamefont {Lee}, \citenamefont {Li}, \citenamefont {Eladl},
  \citenamefont {Zheng}, \citenamefont {Hoffmann},\ and\ \citenamefont
  {Zhu}}]{2404.19135}%
  \BibitemOpen
  \bibfield  {author} {\bibinfo {author} {\bibfnamefont {S.~T.}\ \bibnamefont
  {Chyczewski}}, \bibinfo {author} {\bibfnamefont {H.}~\bibnamefont {Lee}},
  \bibinfo {author} {\bibfnamefont {S.}~\bibnamefont {Li}}, \bibinfo {author}
  {\bibfnamefont {M.}~\bibnamefont {Eladl}}, \bibinfo {author} {\bibfnamefont
  {J.-F.}\ \bibnamefont {Zheng}}, \bibinfo {author} {\bibfnamefont
  {A.}~\bibnamefont {Hoffmann}},\ and\ \bibinfo {author} {\bibfnamefont
  {W.}~\bibnamefont {Zhu}},\ }\href@noop {} {\bibinfo {title} {Strong
  damping-like torques in wafer-scale mote${}_2$ grown by mocvd}} (\bibinfo
  {year} {2024}),\ \Eprint {https://arxiv.org/abs/arXiv:2404.19135}
  {arXiv:2404.19135} \BibitemShut {NoStop}%
\end{thebibliography}%

\end{document}

\title{Supplemental Material for ``Unconventional Spin-Orbit Torques from Sputtered MoTe$_2$ Films"}

\author{Shuchen Li}
\email{sl117@illinois.edu}
\affiliation{%
Department of Materials Science and Engineering and Materials Research Laboratory, University of Illinois Urbana-Champaign, Urbana, Illinois 61801, USA
}%

\author{Jonathan Gibbons}%
\affiliation{%
Department of Materials Science and Engineering and Materials Research Laboratory, University of Illinois Urbana-Champaign, Urbana, Illinois 61801, USA
}%
\affiliation{%
Department of Physics, University of California -- San Diego, La Jolla, California 92093, USA
}%

\author{Stasiu Chyczewski}
\affiliation{Department of Electrical and Computer Engineering, University of Illinois Urbana-Champaign, Urbana, Illinois 61801, USA}

\author{Zetai Liu}
\affiliation{Department of Electrical and Computer Engineering, University of Illinois Urbana-Champaign, Urbana, Illinois 61801, USA}

\author{Hsu-Chih Ni}
\affiliation{%
Department of Materials Science and Engineering and Materials Research Laboratory, University of Illinois Urbana-Champaign, Urbana, Illinois 61801, USA
}%

\author{Jiangchao Qian}
\affiliation{%
Department of Materials Science and Engineering and Materials Research Laboratory, University of Illinois Urbana-Champaign, Urbana, Illinois 61801, USA
}%

\author{Jian-Min Zuo}
\affiliation{%
Department of Materials Science and Engineering and Materials Research Laboratory, University of Illinois Urbana-Champaign, Urbana, Illinois 61801, USA
}%

\author{Jun-Fei Zheng}
\affiliation{Entegris Inc. Danbury, Connecticut 06810}

\author{Wenjuan Zhu}
\affiliation{Department of Electrical and Computer Engineering, University of Illinois Urbana-Champaign, Urbana, Illinois 61801, USA}

\author{Axel Hoffmann}%
 \email{axelh@illinois.edu}
\affiliation{%
Department of Materials Science and Engineering and Materials Research Laboratory, University of Illinois Urbana-Champaign, Urbana, Illinois 61801, USA
}%
\date{\today}

\floatsetup[figure]{style=plain,subcapbesideposition=top}

\maketitle

\tableofcontents

\begin{linenumbers}
\section{Oersted torque efficiencies}
\begin{figure}[htbp]
        \centering
        \includegraphics[width=\columnwidth]{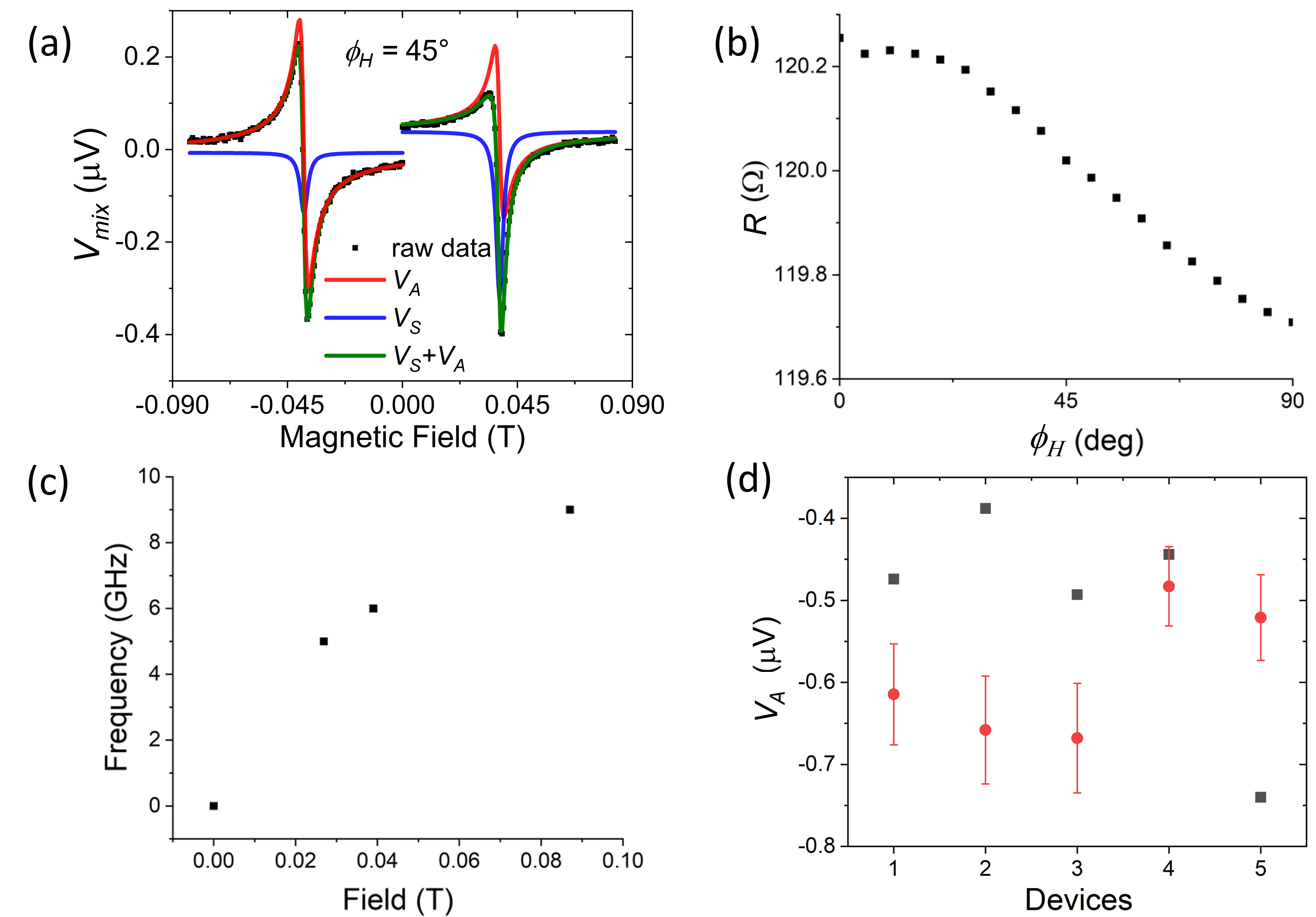}
        \caption{(a) The measured {\em dc} mixing voltages of device 1 from MoTe$_2$(15)/Py/MgO at $\phi_{H}$ = 45$^\circ$ and $\phi_I$ = 0$^\circ$   for positive and negative magnetic field scans. The power and frequency of the current is 4 dBm and 6~GHz. The fit for the mixing voltage is the green curve, which is the sum of symmetric ($V_S$) and antisymmetric ($V_A$) Lorentzians. (b) Anisotropic magnetoresistance of MoTe$_{2}$(15)/Py(10)/MgO as a function of angle $\phi_{H}$. (c) Resonance frequency of the Py at different resonant fields H$_{0}$. (d) Comparison of the calculated antisymmetric voltages contributed by the Oersted field(red) and the experimental results of the extracted antisymmetric voltages(grey) from the ST-FMR signal for different devices from MoTe$_{2}$(15)/Py(10)/MgO at $\phi_{H}$ = 45$^\circ$.}
        \label{S1}
\end{figure}

We performed a two-wire resistance measurement of our 15-nm MoTe$_{2}$ devices and 10-nm Py devices, and obtained the resistivities to be 1333 ${\mu}{\Omega}cm$ and 63.5 ${\mu}{\Omega}cm$ respectively. We have measured the reflection coefficient $S_{11}$ for device 1 of our MoTe$_{2}$(15)/Py(10)/MgO sample and the transmission coefficient $S_{21}$ of our electric circuits under an applied frequency ranging from 0 to 10 GHz using a vector network analyzer (VNA), with $S_{11}$ = -20 dB and $S_{21}$ = -8 dB at 6 GHz. We then used the following equation 
\begin{equation}
    I_{RF} = 2\sqrt{1mW \times 10^{\dfrac{P_{s}+S_{21}}{10}}(1 - 10^{S_{11}/20})^2/50\Omega}
\end{equation}
from Ref~\cite{macneill_control_2017} to calculate the actual current flowing through our devices, where $P_{S}$ is the power output from the signal generator, and obtained the rf current flowing through the device 1 to be 6.25 mA, with 0.417 mA through the MoTe$_{2}$ layer by using a parallel resistor model. Assuming the sample to be an infinitely wide plate, the Oersted field generated by the current flowing through the MoTe$_{2}$ layer can be derived by Ampere’s law $H_{rf} = I_{rf}/2w$, where $w$ is the width of the device and is 42.5 ${\mu}m$ for device 1. 
We then used the formula 
\begin{equation}
    V_{mix} = -\dfrac{1}{4}\dfrac{dR}{d\phi_{H}}\dfrac{{\gamma}I_{rf}cos\phi_{H}}{{\Delta}2{\pi}(df/dH)}[SF_{S}(H_{ext})+AF_{A}(H_{ext})]
\end{equation}
from Ref~\cite{PhysRevLett.106.036601} to calculate the ST-FMR antisymmetric voltage contributions from the Oersted field at $\phi_{H}$ = 45$^\circ$. Here, \textit{R} is the resistance of the device, $\gamma$ is the gyromagnetic ratio, $I_{rf}$ is the rf current through the device, $\Delta$ is the linewidth of the symmetric or antisymmetric signals at the resonance field $H_{0}$, $F_{S}(H_{ext})$ and $F_{A}(H_{ext})$ are the fitted symmetric and antisymmetric Lorentzian functions, $M_{eff}$ = 0.8 T is the effective magnetization of Py, $H_{ext}$ = 0.04 T for the resonance field $H_{0}$ from Fig.~\ref{S1}(a), and $A = H_{rf}[1+(4{\pi}M_{eff}/H_{ext})]^{1/2}$. We performed an AMR measurement for our MoTe$_{2}$(15)/Py/MgO device 1 as shown in Fig.~\ref{S1} (b) and have dR/d$\phi_{H}$ = 0.09 ohm/radian at $\phi_{H}$ = 45$^\circ$. We also have the linewidth $\Delta$ = 0.0017 T at the resonance as shown in Fig.~\ref{S1}(a) and have measured df/dH = 1GHz/0.015T(Fig.~\ref{S1}(c)) for device 1. Thus, we estimated the antisymmetric voltage generated by the Oersted field in the MoTe$_{2}$ layer in device 1 of MoTe$_{2}$(15)/Py/MgO sample is -0.6146 ${\mu}V$ at $\phi_{H}$ = 45$^\circ$, and comparing with the antisymmetric voltages extracted by fitting the Lorentzian component of the ST-FMR signal generated by the device 1 of MoTe$_{2}$(15)/Py/MgO at $\phi_{H}$ = 45$^\circ$, which is -0.474 ${\mu}V$, we think the calculated voltages generated by the Oersted field is within the same scale with the experimental results. We also plot the calculated antisymmetric voltages generated by the Oersted fields (red dots) for several other devices together with the corresponding experimental results (grey dots), which indicates $\xi^{y}$ is dominated by the Oersted field.
\end{linenumbers}

\begin{linenumbers}
\section{Spin-orbit torques due to $x$ spins}
\begin{figure}[htbp]
        \centering
        \includegraphics[width=\columnwidth]{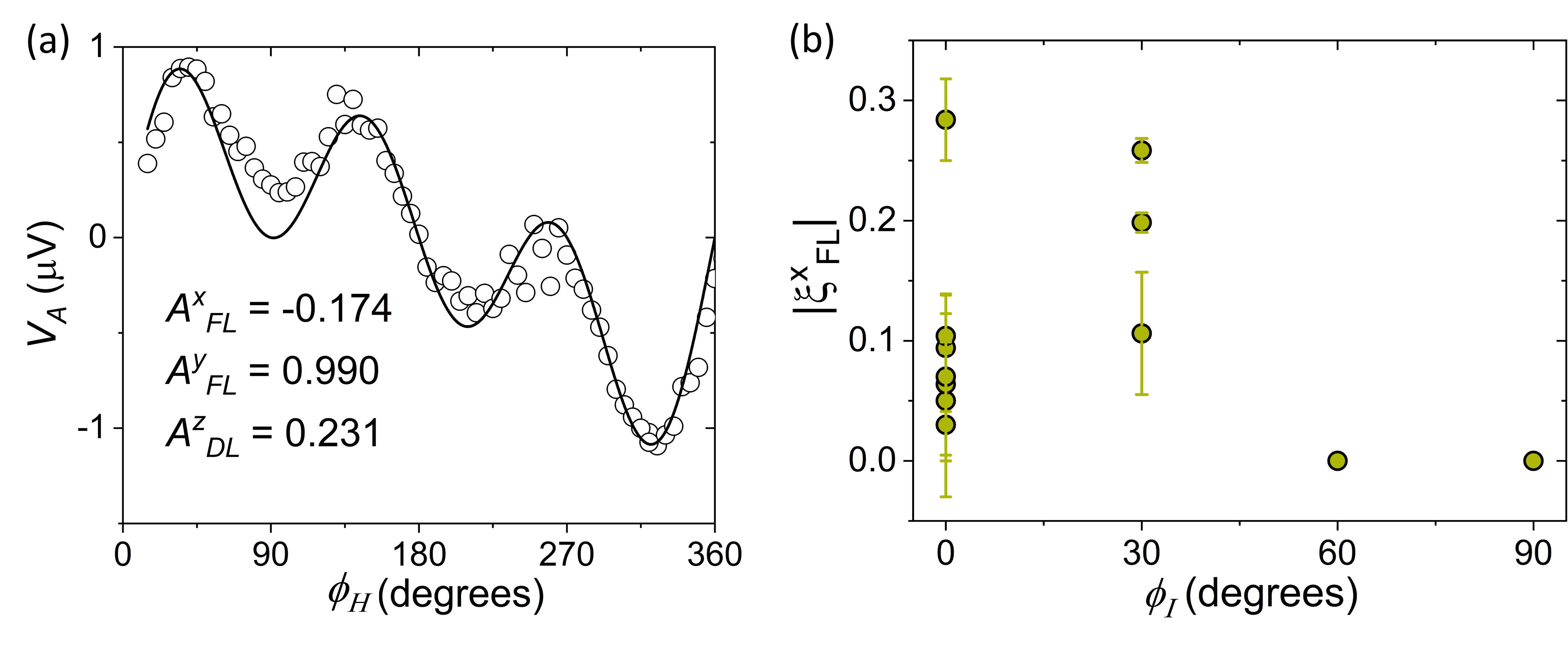}
        \caption{(a) Antisymmetric component(open circles) of a device from MoTe$_2$(15)/Py/MgO with $\phi_{I}$ = 0$^{\circ}$ as a function of angle $\phi_{H}$. The black fit line shows the appearance of $\vec{\tau}{_{FL}^{y}}$, $\vec{\tau}{_{FL}^{x}}$ and $\vec{\tau}{_{DL}^{z}}$. (b) Spin-torque efficiencies $\xi{_{FL}^{x}}$ of devices from MoTe$_2$(15)/Py(10)/MgO(2) samples as a relation to $\phi_{I}$.}
        \label{xspin}
\end{figure}

We also investigated the relation between $\vec{\tau}{_{FL}^{x}}\propto\hat{m}\times\hat{x}$ and $\phi_{I}$ in our MoTe$_2$(15)/Py(10)/MgO(2) devices. We did observe a small $\vec{\tau}{_{FL}^{x}}$ for a few MoTe$_2$(15)/Py(10)/MgO(2) devices when we applied a current parallel to the [$1\bar{1}00$] direction($\phi_{I}$ = 0$^{\circ}$). Fig.~\ref{xspin}(a) shows one device with $\vec{\tau}{_{FL}^{x}}$ in the antisymmetric voltage $V_{A}$ when we applied a current at $\phi_{I}$ = 0$^{\circ}$. However, we also observed a more significant size of $\vec{\tau}{_{FL}^{x}}$ (see Fig.5(a)) for all MoTe$_2$(15)/MgO/Py devices when we applied a current at $\phi_{I}$ = 30$^{\circ}$. The torque efficiency $\xi{_{FL}^{x}}$ of different devices are calculated and plotted as a function of $\phi_{I}$ in Fig.~\ref{xspin}(b). The polarities of the torque $\vec{\tau}{_{FL}^{x}}$ vary from device to device, and we plot the absolute value of $\xi{_{FL}^{x}}$. We noticed that besides one scattered point at $\phi_{I}$ = 0$^{\circ}$, the torque efficiencies $\xi{_{FL}^{x}}$ at $\phi_{I}$ = 0$^{\circ}$ are relatively small compared to those at $\phi_{I}$ =  30$^{\circ}$, not to mention that $\vec{\tau}{_{FL}^{x}}$ are not observed for most devices at $\phi_{I}$ = 0$^{\circ}$. 
\end{linenumbers}

\begin{linenumbers}
\section{Spin-orbit torques from 40 and 7~nm M\MakeLowercase{o}T\MakeLowercase{e}$_2$ samples}

Figure~\ref{40&7STFMR}(a) and (d) consists of the mixing voltages of MoTe$_2$(40)/Py/MgO and MoTe$_2$(7)/Py/MgO samples at $\phi_{H}$ = 315$^\circ$ and $\phi_{I}$ = 0$^\circ$, and exotic torques are also present in both samples as indicated by the distinct lineshapes from positive and negative field scans. To further study different spin-torque components in our sputtered 40-nm and 7-nm MoTe$_2$ samples, similarly, we plot symmetric and antisymmetric components of $V_{mix}$ from MoTe$_2$(40)/Py/MgO and MoTe$_2$(7)/Py/MgO as a function of angle $\phi_{H}$ separately, and as shown in Figure~\ref{40&7STFMR}(c)-(f), we extracted the sizes of different spin-orbit torque components in MoTe$_2$(40)/Py(10)/MgO(2) and MoTe$_2$(7)/Py(10)/MgO(2) respectively. 
\end{linenumbers}
\begin{figure*}[htbp]
        \centering
        \includegraphics[width=\textwidth]{40&7STFMR.png}
        \caption{$V_{mix}$ of (a) MoTe$_2$(40)/Py/MgO and (b) MoTe$_2$(7)/Py/MgO from positive and negative field scans by ST-FMR. We applied a 6-GHz and 9-GHz frequency {\em rf} current through our 40-nm and 7-nm MoTe$_2$ samples respectively. (b) Symmetric and (c) Antisymmetric component of $V_{mix}$ as a function of angle $\phi_{H}$ for MoTe$_2$(40)/Py/MgO. (e) Symmetric and (f) Antisymmetric component of $V_{mix}$ as a function of angle $\phi_{H}$ for MoTe$_2$(7)/Py/MgO. }
        \label{40&7STFMR}
\end{figure*}

\begin{linenumbers}
\section{Comparison with Chemically Deposited $1T'$ M\MakeLowercase{o}T\MakeLowercase{e}$_2$}
\begin{figure*}[htbp]
        \centering
        \includegraphics[width=\textwidth]{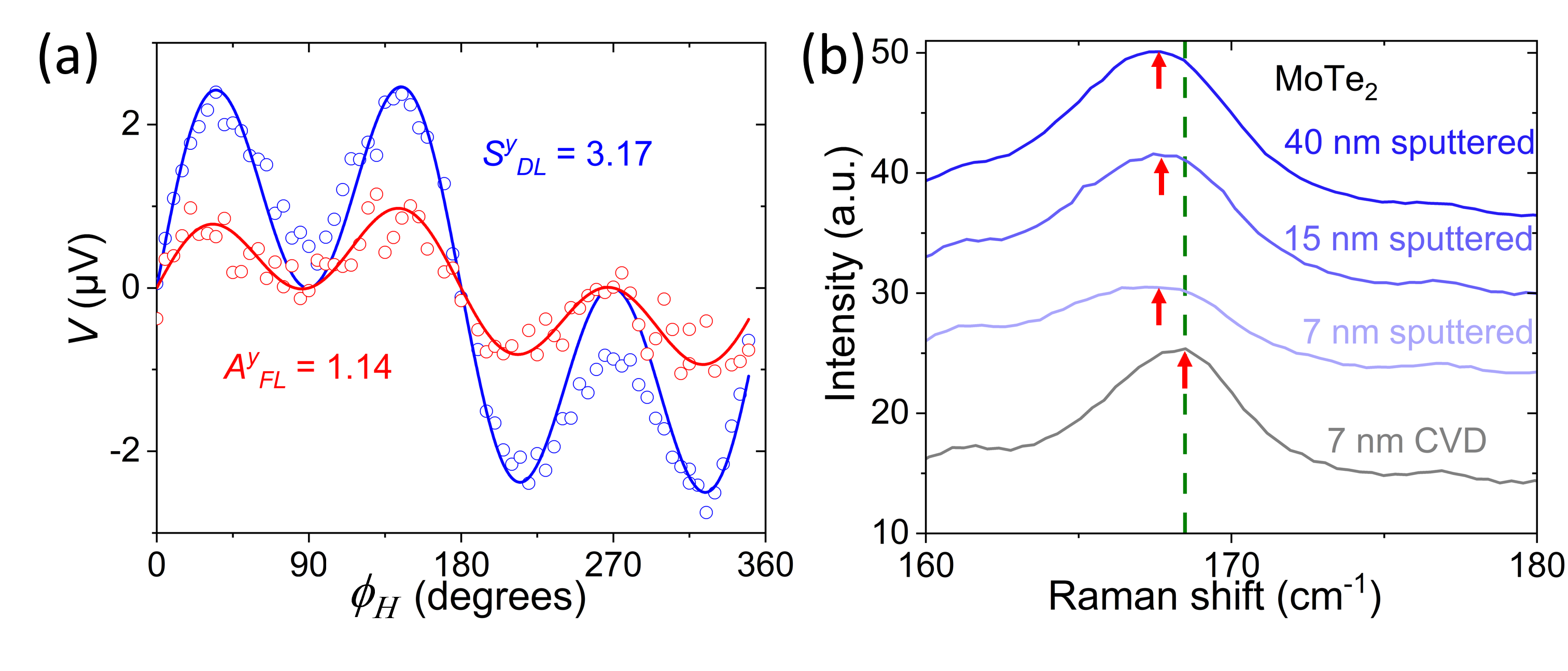}
        \caption{(a) Symmetric $V_S$(blue) and Antisymmetric $V_A$(red) component as a function of $\phi_{H}$ measured for CVD grown films. The red and blue lines are the fit curves for the corresponding raw data. (b) Raman spectroscopies for 40 nm, 15 nm, 7 nm sputtered MoTe$_2$/Py samples, and 7~nm CVD grown MoTe$_2$/Py sample. The laser used is unpolarized with 532 nm exicitation. The red arrows indicate the shifts of the peaks in three sputtered MoTe$_2$ films as compared with CVD grown MoTe$_2$ samples.}
        \label{CVD}
\end{figure*}

We also deposited 10-nm Py and 2-nm MgO layers on 7-nm $1T'$ MoTe$_2$ films grown by chemical vapor deposition(CVD; see Supplemental Material S6 for growth methods), and compared the types of spin-orbit torques generated from the ST-FMR measurements with those found in the sputtered $1T'$ MoTe$_2$ samples, so that we may get more insights on the effects of synthesizing methods on the properties of the MoTe$_2$ films. Similarly, we plot ${S}$ and ${A}$ components as a function of $\phi_{H}$ and fit the angular dependence of ${S}$ and ${A}$ using Eq. (2) and (3) for the CVD-grown MoTe$_2$. As shown in Figs.~\ref{CVD}(a), we only observed contributions from the conventional in-plane damping-like torques $\vec{\tau}{_{DL}^{y}}$ in ${S}$ and the Oersted torques in ${A}$, whereas both $\vec{\tau}{_{DL}^{z}}$ and $\vec{\tau}{_{FL}^{z}}$ due to $z$ spins are not present. We found a fit value of $S{_{DL}^{y}}$ = 0.139 and $A{_{FL}^{y}}$ = 0.096, and have calculated the SOT efficiencies $\xi{_{DL}^{y}}$ from CVD synthesized MoTe$_2$ to be around 0.42, which is similar to that found in the exfoliated MoTe$_2$(0.35) \cite{liang_spin-orbit_2020}.

To better understand the origins of such drastically different spin-torque components in the sputtered and the chemically deposited $1T'$ MoTe$_2$ samples, we compared the Raman spectroscopies of $1T'$ MoTe$_2$ grown by CVD and magnetron-sputtered MoTe$_2$ samples. As shown in Fig.~\ref{CVD}(b), we found that the Raman peaks around 168~cm$^{-1}$ of all sputtered MoTe$_2$ films also shifted to lower energies as compared with those of CVD-grown samples.
\end{linenumbers}

\begin{linenumbers}
\section{Growth of $1T'$ M\MakeLowercase{o}T\MakeLowercase{e}$_2$ film with CVD}
$1T'$ phase MoTe$_2$ was grown on Si/SiO$_2$ substrates using standard chemical vapor deposition techniques \cite{RN460}. A 2 inch, single temperature zone quartz clam shell tube furnace was used. Target substrates were coated with a PTAS solution to promote crystal growth and were then placed in the middle of the heating zone. An MoO$_3$ precursor was placed slightly upstream and a Te powder precursor was placed further upstream such that the temperature of the Te precursor was less than the MoO$_3$ and substrate. Mass ratio of the MoO$_3$ to Te precursors was approximately 1:20. Growth was conducted at atmospheric pressure by first filling the tube with argon before starting the actual process. For growth, the temperature was ramped to 760 $^\circ$ Celsius over the course of 19 minutes and held at that temperature for 21 minutes. The furnace was then allowed to cool naturally before being quenched at 350 degreee Celsius. From the start of the temperature ramp until quenching, 80 sccm of argon and 4 sccm of hydrogen was flowed.
\end{linenumbers}

%old format for figures
%\begin{figure*}[tph]
%\centering \hspace*{\fill} 
%\subfigure[~] {
%\label{fig:S6_a}
%\includegraphics[width=0.3\linewidth]{20210528_supp/UMRvsJ_fit.pdf}} 
%\subfigure[~] {
%\label{fig:S6_b}
%\includegraphics[width=0.3\linewidth]{20210528_supp/linearUMR.pdf}} 
%\subfigure[~] {
%\label{fig:S6_c}
%\includegraphics[width=0.3\linewidth]{20210528_supp/cubicUMR.pdf}} 
%\hspace*{\fill}
%\caption{(a) The current dependence $\alpha J+\beta J^3$ fitted to each applied magnetic field, where T = 10 K and the current line of microwire aligned to FeRh [110]. (b) Extracted linear UMR with respect to the field B. (c) Coefficient $\beta$ for the cubic UMR contribution plotted with respect to the field B.}
%\end{figure*}

%\bibliographystyle{my-aps-style}
%\bibliographystyle{naturemag}
\bibliography{MoTe2.bib}